\documentclass[twocolumn]{aastex6}

\usepackage[T1]{fontenc}
\usepackage{graphicx}
\usepackage{url}
\usepackage{amsmath}
\usepackage[T1]{fontenc}
\usepackage{aecompl}
\usepackage{color}
\usepackage[capitalize]{cleveref}

\graphicspath{{./fig/}}

\crefname{figure}{Fig.}{Figs.}

\newcommand{\vsini}{\ensuremath{v \sin i}}
\newcommand{\Kepler}{\mbox{\textit{Kepler}}}
\newcommand{\Gaia}{\mbox{\textit{Gaia}}}

\newcommand{\Teff}{\ensuremath{T_{\textrm{eff}}}}
\newcommand{\logg}{\ensuremath{\log g}}
\newcommand{\kms}{\textrm{~km~s}\ensuremath{^{-1}}}
\newcommand{\MK}{\ensuremath{M_{Ks}}}
\newcommand{\feh}{\textrm{[Fe/H]}}

\newcommand{\STARBAD}{\texttt{STAR\_BAD}}
\newcommand{\STARWARN}{\texttt{STAR\_WARN}}

\shorttitle{Rapid Rotator Binarity}
\shortauthors{Simonian et al.}
\begin{document}

\title{Rapid Rotation in the \Kepler{} Field: Not a Single Star
Phenomenon}
\author{Gregory V. A. Simonian; Marc H. Pinsonneault; Donald M. Terndrup}
\affil{Department of Astronomy, The Ohio State University \\ 140 West 18th Avenue, Columbus, OH 43210}
\email{simonian@astronomy.ohio-state.edu}

\begin{abstract}
    Tens of thousands of rotation periods have been measured in the
    \Kepler{} fields, including a substantial fraction of rapid rotators. We 
    use \Gaia{} parallaxes to distinguish photometric binaries (PBs) from
    single stars on the unevolved lower main sequence, and compare their
    distribution of rotation properties to those of single stars both with and 
    without APOGEE spectroscopic characterization. We find that 59\% of stars
    with \(1.5 \textrm{ day } < P < 7\) day lie 0.3 mag above the main
    sequence, compared with 28\% of the full rotation sample. The fraction of 
    stars in the same period range is \(1.7 \pm 0.1\%\) of the total sample 
    analyzed for rotation periods. Both the photometric binary fraction
    and the fraction of rapid rotators are consistent with a population of 
    non-eclipsing short period binaries inferred from \Kepler{} eclipsing 
    binary data after correcting for inclination. This suggests that the rapid rotators 
    are dominated by tidally-synchronized binaries rather than single-stars 
    obeying traditional angular momentum evolution. We caution against 
    interpreting rapid rotation in the \Kepler{} field as a signature of youth. 
    Following up this new sample of 217 candidate tidally-synchronized binaries 
    will help further understand tidal processes in stars.
\end{abstract}
\keywords{binaries: close, stars: late-type, stars: rotation}

\section{Introduction}

Rotation is a fundamental property of stars; it impacts their lifetimes, can
induce mixing, and can serve as a population diagnostic. Stars are also born 
with a wide range of rotation rates \citep{Attridge92, Herbst00, Henderson12}.
The observed rotation rates are also strikingly different in high and low mass 
stars. \citet{Kraft67} noted a dichotomy in rotation rates: field stars 
hotter than 6200 K rotate with a wide range of \vsini, while those cooler than 
6200 K, including the Sun, rotate with largely undetectable \vsini. This abrupt transition is thought to
be due to the onset of convective envelopes for lower-mass stars, which 
enables efficient angular momentum loss through magnetized winds 
\citep{Parker58,Weber67}.

The strong dependence of angular momentum loss on rotation \citep{Kawaler88}
leads to the rapid convergence of a wide range of rotation rates to a unique
mass-dependent value; for solar-type stars convergence is nearly complete by
0.5 Gyr \citep{Pinsonneault89}. Rotation is therefore an age indicator for 
stellar populations \citep{Skumanich72}, and correlations between rotation and
age have been used to derive empirical relationships for field populations, 
an approach sometimes referred to as gyrochronology \citep{Barnes07, Mamajek08, Meibom09,
Angus15}. A full theoretical treatment of gyrochronology requires evolutionary 
stellar models that include a variety of physical effects (see \citet{Gallet13} 
for an overview). Models of angular momentum evolution have usually been 
calibrated using rotation periods in star clusters, which have known ages and 
are easily characterized \citep{Krishnamurthi97, Gallet13, Somers17}. The 
calibrating clusters are typically young (\(< 1\) Gyr), with the behavior at old 
ages anchored by the Sun. 

The \Kepler{} satellite \citep{Borucki10,Koch10}, which observed a single field
for over four years, has revolutionized our ability to measure stellar 
rotation in old field populations. Large-scale analyses have extracted 
rotation periods from the light curves of tens of thousands of field stars
\citep{Nielsen13, Reinhold13, Garcia14, McQuillan14}. These rotation periods, 
along with a field gyrochronology, are expected to provide insights on the age 
distribution of the field, and thus provide a more accurate representation of 
transiting exoplanets.

Recent observations of old field stars enabled by \Kepler{} have challenged 
our theories of angular momentum evolution on multiple fronts. Existing 
gyrochronological models anchored at the Sun have not been able to explain
the relatively rapid rotation of old field stars with ages inferred from
asteroseismic data \citep{Angus15, VanSaders16}. The \Kepler{} population at 
long periods shows a sharp drop-off which is not predicted by standard angular
momentum evolution models. The data appears to require either a dramatic
decrease in magnetized winds, detectability, or a combination of both 
\citep{VanSaders18}.  Nearby K- and M-dwarfs show a bimodality in 
the period distribution which has no clear explanation in stellar physics 
\citep{Davenport18}. There is also a modest, but real, population of rapid 
rotators that would be interpreted as young if they were single main sequence 
stars.

The \Kepler{} field also contains a mixture of stellar populations, not all of 
which have been calibrated in clusters. Subgiants, while being rare in the 
clusters used to calibrate gyrochronology, 
make up 20\% of \Kepler{} targets \citep{Berger18b}. Since the 
subgiant population consists of stars undergoing substantial structural changes
as they expand off the main sequence, they obey a different gyrochronology 
relation than that calibrated for cluster dwarfs \citep{vanSaders13}. 

Another population which is not included in traditional angular momentum
evolution models is tidally-synchronized binaries. These binaries have orbital 
periods short enough that tidal interactions force the rotational period to
synchronize to the orbital period. Existing models of tidal theory 
\citep{Zahn77} predict a strong dependence of the synchronization timescale 
on period, leading to rapid transition between
synchronized and unsynchronized systems. Understanding the 
age distribution of the \Kepler{} field at the young end will require 
characterizing the background population of tidally-synchronized binaries.

Tidally-synchronized binaries themselves are in the center of interesting
dynamical phenomena. These systems are thought to be dynamically-formed 
through three-body interactions \citep{Tokovinin06, Fabrycky07}. Once formed, they can experience significant
evolution in orbital period due to angular momentum loss \citep{Andronov06}.
Depending on the rate of angular momentum loss, they can merge to form blue 
stragglers or pre-CV systems.

Tidally-synchronized binaries have been difficult to study because they are 
intrinsically rare and usually require significant spectroscopic resources 
to discover and characterize \citep{Mathieu90, Raghavan10, Geller15}. A review 
of the observational state of tidal interactions is given in \citet{Mazeh08}. 
Most observations efforts to understand stellar tides have focused on 
measuring the ``cut-off period'' for orbital circularization \citep{Mayor84},
which usually requires intense spectroscopic monitoring to derive orbits.

The largest homogeneous study of synchronicity to date measured starspot 
variability in the \Kepler{} eclipsing binary sample \citep{Lurie17}, which 
found a synchronization period cutoff at 10 days, in agreement with previous studies. 
A novel result found in their sample was the existence of a population of
subsynchronous rotators (at the level of 15\%) with orbital periods between 
2--10 days. Follow-up observations and a larger sample of these binaries may 
provide unexpected insights in the theory of three-body interactions.

The availability of \Gaia{} DR2 parallaxes \citep{Gaia18,Lindegren18} 
enables us to better characterize sources of rapid rotation beyond what is
achievable through photometry and spectroscopy. Young single stars would be
observed as rapid rotators within the field main sequence. Subgiants would lie 
in the Hertzprung Gap, on the way to the Red Giant Branch. Lastly, binaries 
should be up to twice as luminous as the main sequence, extending up to 0.75 
mag above the single-star sequence.  While these populations overlap for hot 
stars, the subgiants and binaries separate cleanly on the lower main sequence.

An additional constraint on the populations comes from the substantial sample
of spectroscopically-characterized dwarfs from the Apache Point Observatory
Galactic Evolution Experiment (APOGEE) \citep{Majewski17}. APOGEE contains a 
total of 15,724 objects in the \Kepler{} field with usable stellar parameters,
\(Ks\)-band photometry, and parallaxes.  This sample will be valuable for 
understanding the importance of metallicity to identify populations of 
binaries.

This paper will attempt to compare binarity properties between the rapid and 
slow rotators in \citet{McQuillan14}. In Section~\ref{sec:data}, we will describe our 
sample, as well as the data used to select binaries. In Section~\ref{sec:analysis}, we demonstrate the regime where 
binaries can be successfully distinguished from single stars with both the 
presence and absence of metallicity information. We'll also characterize the 
uncertainty in our measurements. Section~\ref{sec:results} illustrates our 
results regarding the prominence of binaries among the rapid rotators. Lastly, 
we discuss the implications of our results for the population of rapid rotators 
and lay out future avenues to explore.

\section{Catalog Data}
\label{sec:data}

The base sample of this study consists of \Kepler{} dwarfs that were analyzed
for photometric starspot modulations. All targets in the \Kepler{} field have 
stellar parameters determined at the very least by KIC photometry 
\citep{Brown11}. Based on the KIC photometry, \citet{McQuillan14} selected 
133,030 targets to search for starspot variability, of which 34,030 have 
period detections.

We also examine a spectroscopically-characterized subsample from APOGEE\@. 
This sample has spectroscopically-determined temperatures and metallicities. 
The overlap between this sample and the number of objects inspected by 
\citet{McQuillan14} for rotational period modulation is 3,023.

To calculate vertical displacements above the main sequence, we use photometry
from 2MASS \citep{Skrutskie06}, parallaxes from \Gaia{} DR2 \citep{Gaia18},
\Teff{} from \citet{Pinsonneault12}, extinction values estimated from the 
\Kepler{} Stellar Parameter Catalog (KSPC) DR25 \citep{Huber14,Mathur17}, and
MIST isochrones \citep{Choi16} (described further in
Section~\ref{sec:analysis}). For the spectroscopic sample, \Teff{} and
\feh{} are taken from APOGEE DR14 \citep{Abolfathi18}. More detail on these 
catalogs is given below.

\subsection{Default Stellar Parameters}

Stellar parameters for the full \Kepler{} sample are available as part of the
\Kepler{} Stellar Parameter Catalog (KSPC) DR25 \citep{Mathur17}. The KSPC
compiles stellar parameters from the literature and simultaneously fits them
using DSEP stellar evolutionary tracks \citep{Dotter08} via the methodology of
\citet{Huber14}. Revised stellar parameters, such as \Teff, as well as 
inferred quantities, such as extinction, are output along with asymmetric
1-\(\sigma\) uncertainties. 

The sophisticated machinery of the KSPC imprints artifacts in the \Teff{}
distribution of cool dwarfs by attempting to stitch together the temperature
scales of \citet{Pinsonneault12} and \citet{Dressing13}. Instead of using
\Teff{} from the KSPC, we opted to use temperatures from 
\citet{Pinsonneault12} to ensure uniformly analyzed temperatures; this data is
close to the KSPC scale globally \citep{Huber17}. We note that 
\citet{Pinsonneault12} only provides temperatures for colors where the
infrared-flux method has been well-studied, which corresponds to between
4000--7500 K.

For this paper, we make use of extinctions from the KSPC\@. The
extinctions are inferred from the predicted absolute magnitude given KSPC 
parameters, and a 3D extinction map \citep{Amores05}. The catalog extinction 
is given as \(A_V\), which we convert to other bands using the 
\citet{Cardelli89} extinction law. 

\subsection{Astrometric Data}

The ability to distinguish between different evolutionary states of stars is
enabled by parallaxes derived from the \Gaia{} mission's Data Release 2
\citep{Gaia18}. The \Gaia{} DR2 performed a fully consistent single-star
5-parameter (\(\alpha\), \(\delta\), \(\mu_\alpha\), \(\mu_\delta\), \(\pi\))
solution to 1.3 billion sources over 22 months of observations
\citep{Lindegren18}. For targets which could not be adequately fit with a
single-star 5-parameter solution, a 2-parameter (\(\alpha\), \(\delta\))
solution is performed instead \citep{Michalik15}. The criteria for a
2-parameter fall-back solution include targets with Gaia \(G > 21.0\) mag, 
having fewer than 6 visibility periods of observations, and an error ellipse 
larger than a magnitude-dependent limit in its largest dimension 
\citep{Lindegren18}.

We use the cross-matched database of \citet{Berger18b} to match \Kepler{} targets
against \Gaia{} DR2 detections.
\citet{Berger18b} cross-matched targets in the KSPC DR25 \citep{Mathur17}, 
with \Gaia{} DR2 sources using both position and \(G\)-band flux. Of the 
successfully cross-matched targets, \citet{Berger18b} excluded objects with 
fractional parallax errors greater than 0.2, \(\Teff < 3000\) K, 
\(\logg < 0.1\), and low quality 2MASS photometry.  Because the remaining
targets have high-quality parallaxes, we use the traditional 
formula for deriving distance modulus from parallax instead of the Bayesian 
method advocated by \citet{Luri18} to reduce computational complexity. We
include the zero-point offset of 0.05 mas \citep{Zinn18}.

\subsection{Photometric Data}

In order to characterize our stellar sample, we chose to use absolute 
\(Ks\)-band magnitude as a proxy for luminosity to minimize the impact of 
extinction.  \(Ks\)-band apparent magnitudes were measured by the 2MASS survey 
\citep{Skrutskie06}, which performed profile-fit photometry and derived 
uncertainties for nearly the full sample.  The typical \(Ks\)-band 
photometric uncertainty for the \citet{McQuillan14} sample is 0.03 mag. The 
typical extinction from the KSPC and its uncertainty for cool dwarfs is 
\(A_V = 0.33\), and \(\sigma_{A_V}=0.025\). Applying the \citet{Cardelli89} 
relation that \(A_K/A_V = 0.114\), the effect of extinction itself shrinks to 
the level of photometric errors.  A small fraction of targets (1.5\%) of the 
\citet{McQuillan14} sample have 2MASS photometry flagged due to blending. To 
avoid contaminating the binary sequence with unrelated blends, we exclude 
these targets.

\subsection{Rotation Data}

The rotation periods for our sample come from the catalog compiled by 
\citet{McQuillan14}, which is large and homogeneously measured. 
\citet{McQuillan14} selected their sample 
to have photometric \(\Teff < 6500\) K \citep{Brown11,Dressing13}, and 
implemented the color-color and \logg{} cuts advocated in \citet{Ciardi11}, 
ranging from \(\logg = 3.5\) at 6000 K to \(\logg = 4.0\) at 4250 K. 
In addition to the temperature and gravity cuts, \citet{McQuillan14} also 
excluded known \Kepler{} eclipsing binaries and \Kepler{} Objects of Interest. 

Rotational periods in \citet{McQuillan14} were measured by calculating the 
autocorrelation function (ACF) and fitting the location of multiple peaks.
Pre-processed light curves were taken the PDC-MAP pipeline
\citep{Smith12,Stumpe12}; the light curves were median-normalized and zeroed
in inter-quarter gaps before calculating the ACF \citep{McQuillan13}. In
lieu of visual verification of the autocorrelation function, \citet{McQuillan14}
performed two automated tests to distinguish physical periodicity from
instrumental artifacts or other sources of variability. First, the 
periodicity must be consistent in different segments of the light curve.
Secondly, the height of the first peak must be larger than a temperature- and
period-dependent threshold. In order to reduce contamination from pulsators, 
\citet{McQuillan14} only considered periods between 0.2--70 days.

\begin{figure*}[htb]
    \centering
    \epsscale{1.1}
    \plotone{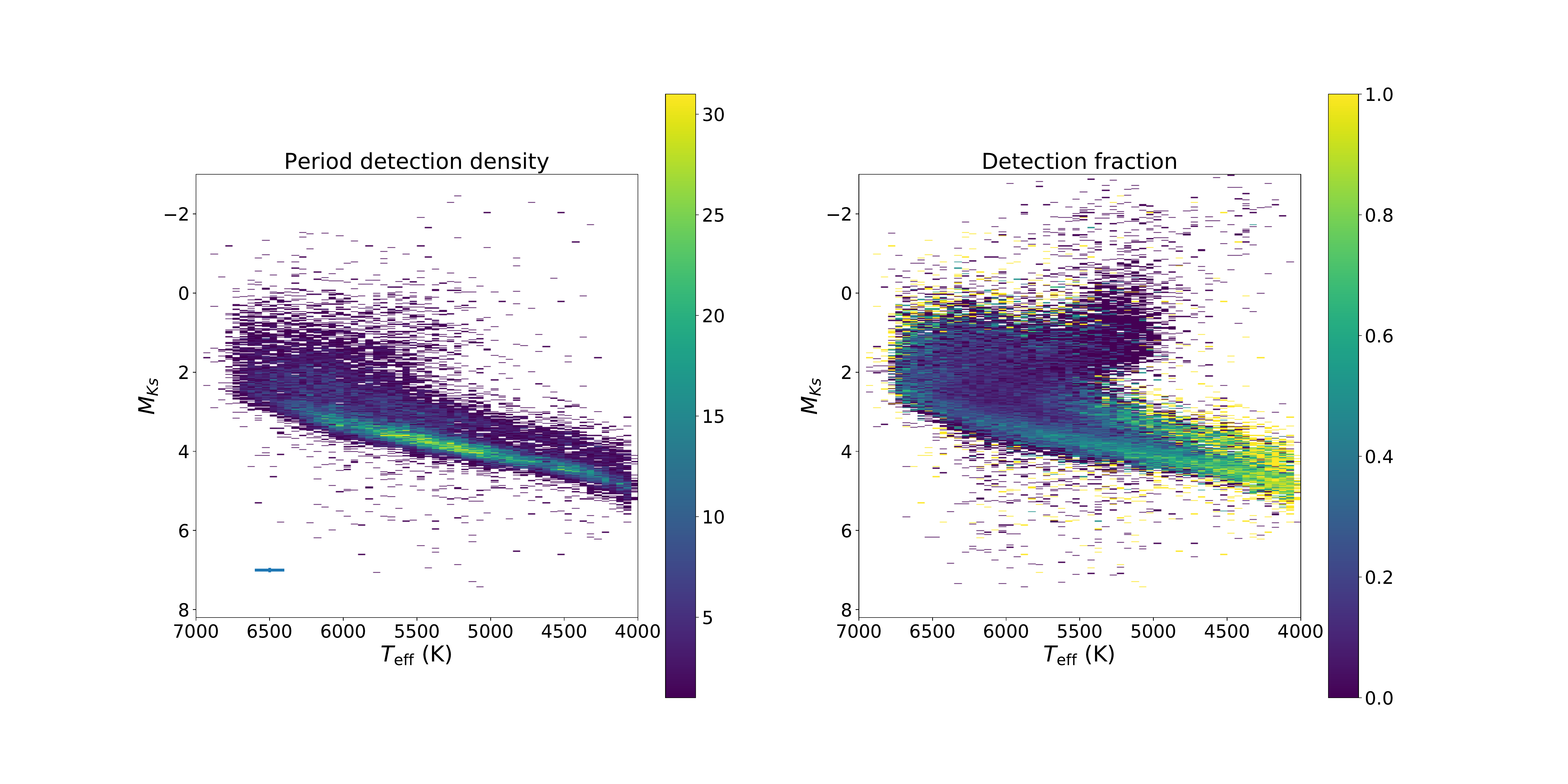}
    \caption{\emph{Left:} \Teff{}-\MK{} density plot of the sample of
        \citet{McQuillan14} period detections. Color represents the number of
        objects in each bin. A binary sequence is clearly 
        visible above the lower main sequence. Temperatures are from
        \citet{Pinsonneault12}. The bin size is 100 K in
        temperature and 0.02 mag in \(Ks\)-band absolute magnitude. A 
        representative error bar is shown on the bottom left corner, although
        the vertical error bar is too small to be easily visible. \emph{Right:} 
        The variation in the \citet{McQuillan14} period 
        detection fraction across the \Teff-\MK{} 
    diagram.}\label{fig:mcquillan_selection}
\end{figure*}

The \citet{McQuillan14} sample is shown in a \Teff-\MK{}
diagram in \cref{fig:mcquillan_selection}. One of the new results enabled by 
\Gaia{} is that a binary sequence,
located above the main sequence, is clearly seen. Also shown in
\cref{fig:mcquillan_selection} is the fraction of stars analyzed within each
bin with period detections. The expected trends in activity are seen: period 
detections are common on the lower main sequence, and become relatively rare 
in evolved stars. The bulk of the period sample are solar analogs, reflecting 
the underlying \Kepler{} sample.

\subsection{Spectroscopic Parameters}
\label{sec:speccat}

We draw our spectroscopic sample from the Data Release 14 \citep{Abolfathi18}
of the APOGEE survey \citep{Majewski17}. The workhorse behind the survey is the
APOGEE spectrograph, a high-resolution (\(R \sim 22,000\)) multi-fiber
near-infrared spectrograph \citep{Wilson10} mounted on the SDSS 2.5-meter
telescope at Apache Point Observatory \citep{Gunn06}.

Reduction of the APOGEE data takes place in three main stages. First,
individual visit spectra are reduced, including detector calibration, bad pixel
masking, wavelength calibration, sky subtraction, and determination of
individual radial velocities through cross-correlation of template spectra.
Second, the individual spectra are combined by correcting for radial velocity
differences between the exposures, either by cross-correlating with each other,
or with template spectra, whichever leads to a smaller scatter
\citep{Holtzman18}. The full process is detailed in \citet{Nidever15}.

The final step is the extraction of stellar parameters and chemical abundances
by the APOGEE Stellar Parameter and Chemical Abundances Pipeline (ASPCAP)
\citep{GarciaPerez16}. ASPCAP measures stellar parameters by performing a
chi-squared minimization \citep{AllendePrieto06} over a 6-dimensional space of
\Teff, \logg, [M/H], [\(\alpha\)/M], \vsini, and microturbulent velocity. Once
those six parameters have been determined, ASPCAP measures abundances for
individual elements, including \feh{}.

After determination of the stellar parameters and chemical abundances, the 
effective temperatures are calibrated to the photometric system of \citet{GonzalezHernandez09} using a 
metallicity-dependent offset, and the abundances are calibrated to provide
homogeneous results within clusters. After calibration, the scatter between the 
calibrated and photometric temperatures  were found to be a function of 
\Teff{}, ranging from 130 K at 5500 K down to 85 K at 4000 K
\citep{Holtzman18}. The statistical uncertainty in \feh{} was measured to be 
0.009 dex \citep{Holtzman18}, but the systematic uncertainty is likely closer
to 0.1 dex \citep{Serenelli17}.

We inspected the fits to a set of spectra representative of targets flagged 
by the ASPCAP pipeline as having potential quality problems. A small fraction
(2.5\%) of targets have the \STARBAD{} quality flag enabled. Visual inspection
of a representative sample of spectra indicated that the flag was largely the
result of poor subtraction or normalization of the spectrum in the pipeline, or
due to poor model fits on the cool (\(\Teff < 4250 \textrm{ K }\)) end. We exclude these
from our analysis. A substantially larger fraction of targets (14\%) have the 
\STARWARN{} quality flag enabled. Visual inspection of a representative sample 
of these spectra indicated that the fits reasonably resembled the underlying 
spectra. We include all targets with the \STARWARN{} quality flag enabled.

\begin{figure*}[htb]
    \centering
    \epsscale{1.1}
    \plotone{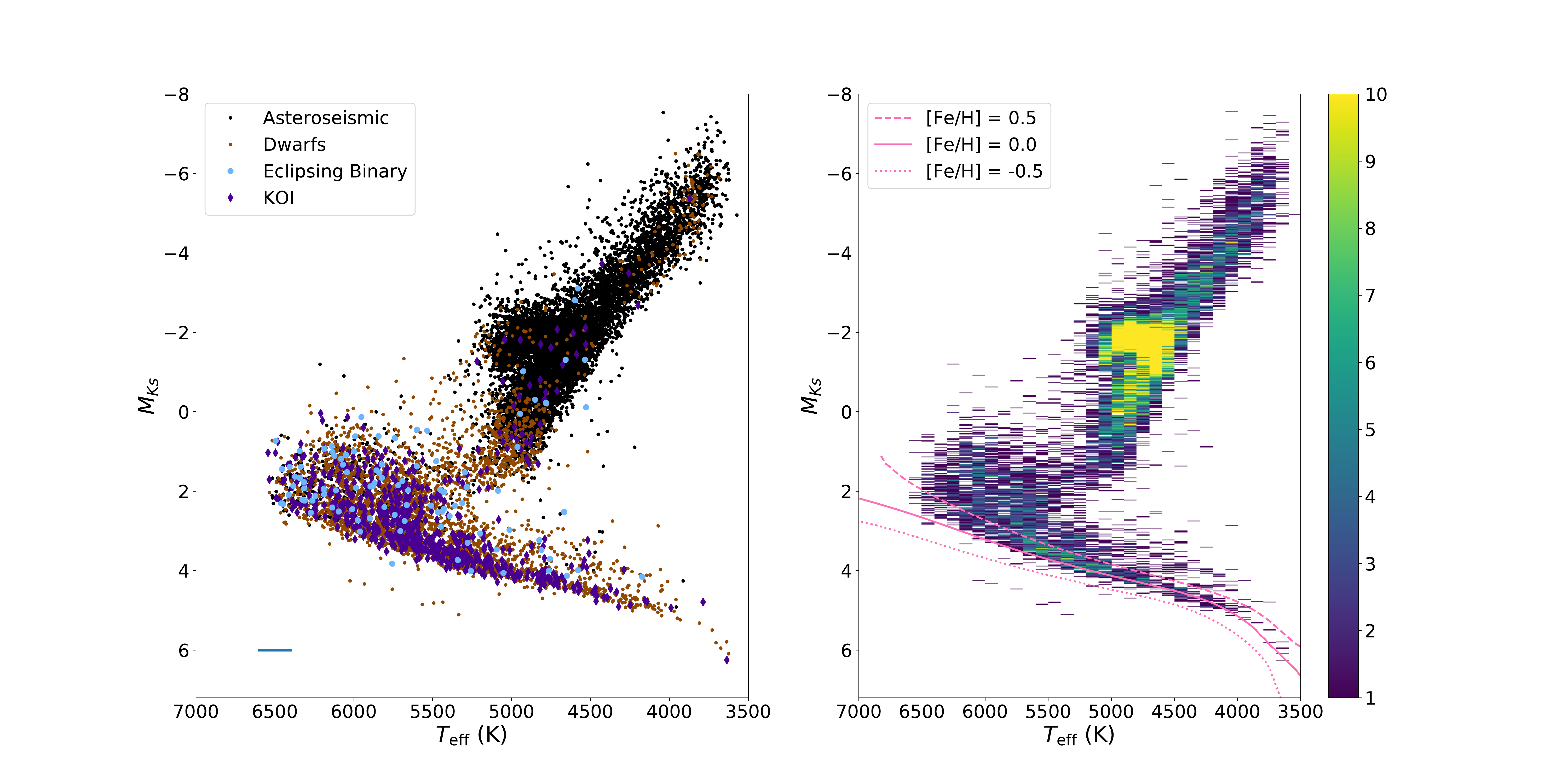}
    \caption{\emph{Left:} \Teff-\MK{} diagram for the APOGEE observations of
        \Kepler{} targets. Asteroseismic targets are shown as black dots. The
        dwarf sample is shown as brown dots. The light blue circles indicate 
        eclipsing binary targets and purple dots are \Kepler{} Objects of 
        Interest. A binary sequence is clearly visible on the lower main
        sequence. Temperatures are spectroscopic APOGEE temperatures. A 
        representative error bar for the cool dwarf sample is 
        shown in the bottom-left corner, although the vertical error bar is too
        small to be easily visible. \emph{Right:} A density plot of the 
        full APOGEE sample. To preserve the dynamic range of the dwarf 
        sequence, the red clump was allowed to saturate. The bin size is 100 K
    in temperature and 0.02 mag in \(Ks\)-band absolute 
magnitude. 1~Gyr \mbox{MIST} isochrones with \([Fe/H] = 0.5 \textrm{ (dashed)}, 
    0.0 \textrm{ (solid), and } -0.5 \textrm{ (dotted)}\) are displayed in 
pink.}\label{fig:apogee_selection}
\end{figure*}

\Cref{fig:apogee_selection} shows all \Kepler{} targets with observed
spectroscopic temperatures in the APOGEE DR14, color-coded by observing
program. One of the driving efforts for \Kepler{} field observations has been
the spectroscopic characterization of asteroseismic targets, shown as small
black dots \citep{Zasowski17,Pinsonneault18}; 11,734 targets are part of that
sample. The asteroseismic sample is substantially biased toward giants 
and subgiants due to their high amplitude oscillations. A complementary 
program proposed to observe the remaining portion of the \Kepler{} sample 
cooler than 6500 K and brighter than \(H < 11\), make up an additional 3193
targets, shown as small red dots. An additional 695 targets were part of an 
effort to characterize \Kepler{} planet hosts with \(H < 14\), shown as purple 
diamonds. The remaining 102 targets were targeted to follow-up 
eclipsing binaries \citep{Prsa11,Slawson11}, and are shown as light blue 
circles.  

The bulk (\(\sim80\%\)) of the cool dwarfs (\(\Teff < 5250\) K were observed through the cool dwarf
programs. These were large magnitude-limited surveys targeting objects with
photometric \(\Teff < 6500\) K, photometric \(\logg \le 3.5\), and 
\(H > 11\)~mag. This is expected to be our least biased sample because there are 
only course astrophysical constraints on the parameters. Much of the remaining 
sample (\(\sim 17\%\)) consists of follow-up observations to \Kepler{} Objects of 
Interest. These observations go much deeper, down to \(H = 14\)~mag. The
inferred presence of a planet implies that this sample is 
likely to be biased toward high-metallicity and single stars. Lastly, the
eclipsing binaries make up the rest of the sample, whose only criterion is
that eclipses are found in the \Kepler{} light curves. Because we do not want
to bias our sample against binaries, include them even though they are few in
number.

\subsection{Eclipsing Binaries}
\label{sec:ebcat}

When selecting the sample to perform their periodogram analysis,
\citet{McQuillan14} excluded known eclipsing binary targets to avoid confusing
the eclipse signal as a rotation signal, which biases the rotation sample
against close binaries. To correct for this bias, we include the 
 \Kepler{} Eclipsing Binary sample \citep{Prsa11,Kirk16}, with the assumption
 that the short-period eclipsing binaries are synchronized.

With \Kepler{}'s continuous monitoring and a long baseline of 4 years, the 
eclipsing binary sample for orbital periods on the order of 20 days should be 
complete \citep{Kirk16}. Most eclipses were identified by the main
\Kepler{} pipeline \citep{Jenkins10}. Orbital periods were measured using a
neural network trained on synthetic eclipsing binary data \citep{Prsa08}. 
We downloaded the eclipsing binary catalog V3 from the Villanova web 
site\footnote{\url{http://keplerebs.villanova.edu/}}, on June 3, 2016. 

\section{Data Analysis}
\label{sec:analysis}

The fundamental classification in this analysis is to distinguish photometric
binaries from the rest of the sample. The crucial quantity to make this
distinction is the 
vertical displacement above the single-star sequence. Vertical 
displacement is a standard measure of binarity in clusters 
\citep{Mermilliod92}, where the single-star sequence can be described by a
stellar population with a unique age and metallicity. Equal-mass binaries
would lie 0.75 mag above the single-star sequence, while equal-mass triples would be
1.25 mag above. Measuring the vertical displacement for a field population is 
more challenging, because the
field consists of heterogeneous ages and metallicities. There is a
field turn-off, however, and for sufficiently cool dwarfs there is a
well-defined unevolved main sequence that is an analog of the cluster case. We
therefore search for the temperature domain where age effects are minimized, 
and where the natural width of the unevolved main sequence can be confidently
measured to identify field binaries.

The vertical displacement is a difference of two quantities: the measured 
\(Ks\)-band luminosity of a star, and the inferred \(Ks\)-band luminosity of a 
single star with the same temperature and metallicity as the original one at a 
reference age. The former is a well-constrained and easily calculated quantity 
given the 2MASS photometry and the \Gaia{} parallax; the latter requires a 
single-star model.

\subsection{Stellar Model}

We use the MIST \citep{Dotter16,Choi16} isochrones to model the 
single-star lower main sequence. Bolometric corrections translating from
evolutionary tracks to photometric bands are generated from ATLAS12/SYNTHE
stellar atmosphere models \citep{Kurucz70,Kurucz93}. The solar abundance scale
for the isochrones is from \citet{Asplund09}. For a full description of the 
MIST isochrones, we refer the reader to the source papers, as well as the MESA 
instrument papers \citep{Paxton11, Paxton13, Paxton15}, which is the stellar 
model on which the isochrones are based. In order to characterize the vertical 
displacement, we select 1 Gyr isochrones
as a baseline case. A sample 1~Gyr solar metallicity isochrone is shown in
\cref{fig:apogee_selection}. As age effects in the domain of interest are 
by definition small, the precise choice of baseline age does not significantly impact our 
results (see Section~\ref{sec:age}).

\subsection{Isolating Unevolved Stars}
\label{sec:age}

\begin{figure}[htb]
    \centering
    \plotone{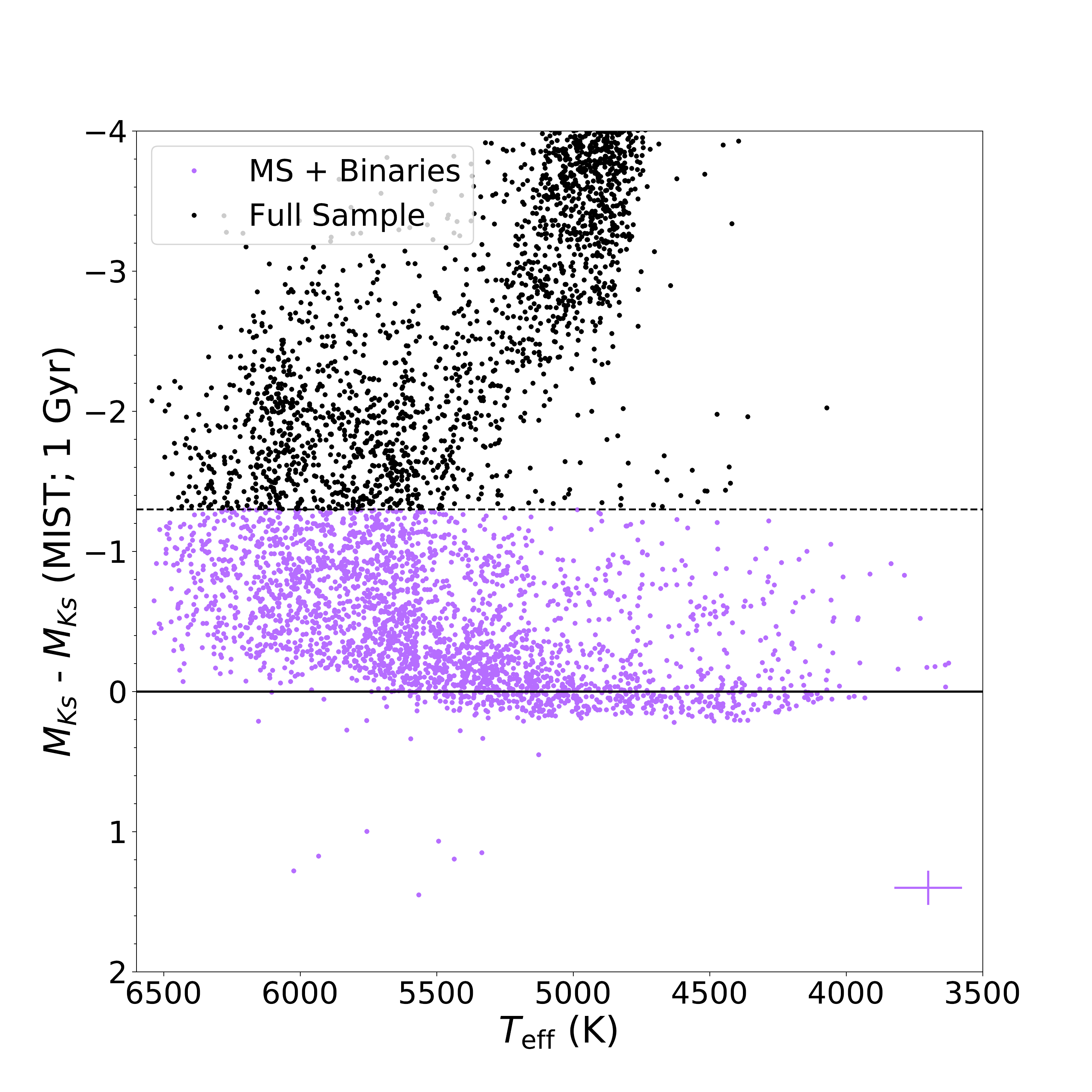}
    \caption{APOGEE temperature and vertical displacement above a 1 Gyr MIST 
        isochrone matched to the APOGEE metallicity for the APOGEE-\Kepler{} 
        sample.  In order to exclude evolved stars while preserving binaries 
        and triples, we define the dwarf sample as targets less than 1.3 
        magnitudes more luminous than the metallicity-adjusted MIST isochrone 
        (shown as violet points). A representative error bar for the 
        full APOGEE dwarf sample assuming Gaussian uncertainties is shown in 
        the lower right corner.}\label{fig:sample_dk}
\end{figure}

The effect of age is degenerate with that of binarity on the main
sequence. Older stars at a given temperature are more luminous than younger 
stars. However, low mass stars experience little luminosity change over the 
lifetime of the Milky Way galaxy, therefore age will have a minimal impact on 
their apparent luminosity. As a result, the unevolved lower main sequence is an 
ideal regime for finding binaries. The transition where the age effect becomes 
small can be clearly measured by calculating  the vertical displacement of the 
full APOGEE sample above a 1 Gyr isochrone, which is shown in 
\cref{fig:sample_dk}. 

For every star, the displacement is calculated by subtracting the luminosity 
of the 1 Gyr isochrone with the APOGEE-measured 
metallicity. F and G stars spread above the isochrone, which is 
expected because these stars evolve substantially over the age of the Milky
Way. With the main sequence detrended, a visible binary sequence emerges at 
\(\sim0.75\) mag above the main sequence. Substantially above the main sequnce
are the red giant branch stars, which we exclude by drawing a limit at 1.3 
magnitudes above the main sequence. This limit cleanly excludes red giant 
branch stars while providing generous inclusion limits for binaries and 
potential triple systems.

\begin{figure}[htb]
    \centering
    \plotone{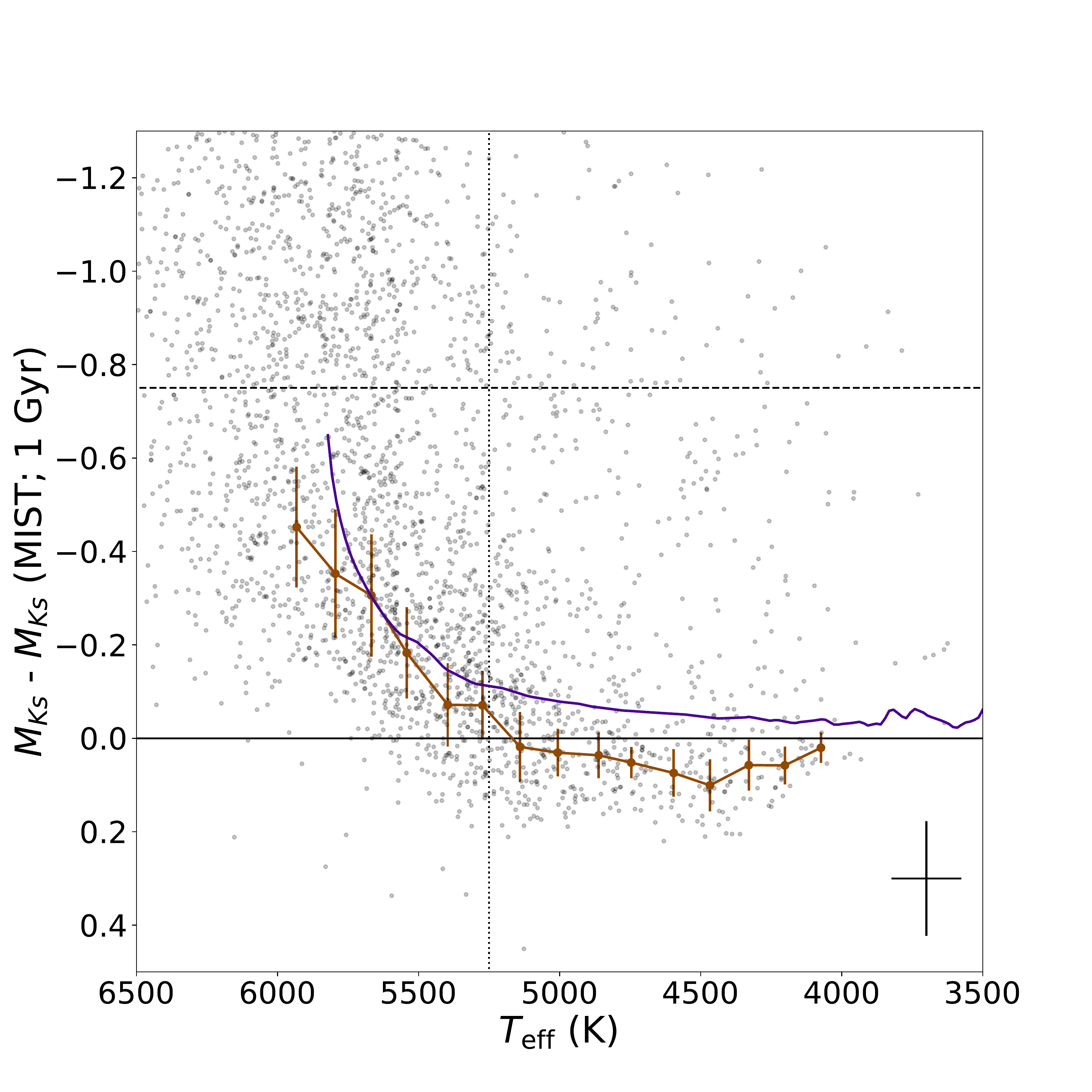}
    \caption{Age-imposed behavior for the dwarf sample with APOGEE
        temperatures. The purple line denotes the predicted luminosity excess of 
        a 9 Gyr solar composition isochrone relative to a 1 Gyr baseline at
        fixed metallicity. The dashed line represents 0.75 
        mag, the limit for photometric binaries. The 25th percentile of the 
        dwarfs in 15 temperature bins is shown in brown. The median absolute 
        deviation of the least-luminous 50\% of the bin is shown as the error 
        bar. We draw a temperature threshold for the unevolved lower main 
        sequence at 5250 K. A representative statistical error bar assuming 
        Gaussian uncertainties is shown in the lower right corner.}
    \label{fig:ages}
\end{figure}

To reduce the impact of age,
we restrict ourselves to the unevolved lower main sequence, where age
effects are small. We quantify the impact of age by predicting the vertical
displacement of a 9~Gyr solar-composition MIST isochrone with respect to our 
1 Gyr solar-composition MIST isochrone, shown in \cref{fig:ages}. The
predicted displacement between the two isochrones is -0.20~mag at 5500~K, 
-0.11~mag at 5250~K, and -0.08~mag at 5000~K, showing a natural break at 
5250~K.

We parametrize the behavior of the main sequence by splitting it into bins
and calculating the 25th percentile of the luminosity excess in each bin. Each 
temperature bin contains a population of both binaries and single stars, with 
binaries having substantially higher vertical displacements. The 25th 
percentile statistic characterizes the main sequence in a way that is robust 
to the population of binaries, and has been used successfully in clusters 
\citep{An06}. Even large departures from the assumed binary fraction should 
generally only affect the zero-point, but not the shape of the main sequence.

The detrended main sequence is generally flat at temperatures cooler than 5000 K 
with age effects becoming more important at hotter temperatures. The number of stars
also increases substantially at temperatures above 5000 K. As a compromise
between age effects and statistical power, we adopt a threshold \Teff{} of 
5250 K to denote the lower main sequence.

\subsection{Metallicity Distribution}

For the full \citet{McQuillan14} sample, where individual metallicities are not
available, we need to select an isochrone with a metallicity that is 
representative of the full sample. We do this by assuming that the metallicity 
distribution of the spectroscopic targets is representative of the full 
\Kepler{} field. 

\begin{figure*}[htb]
    \centering
    \plotone{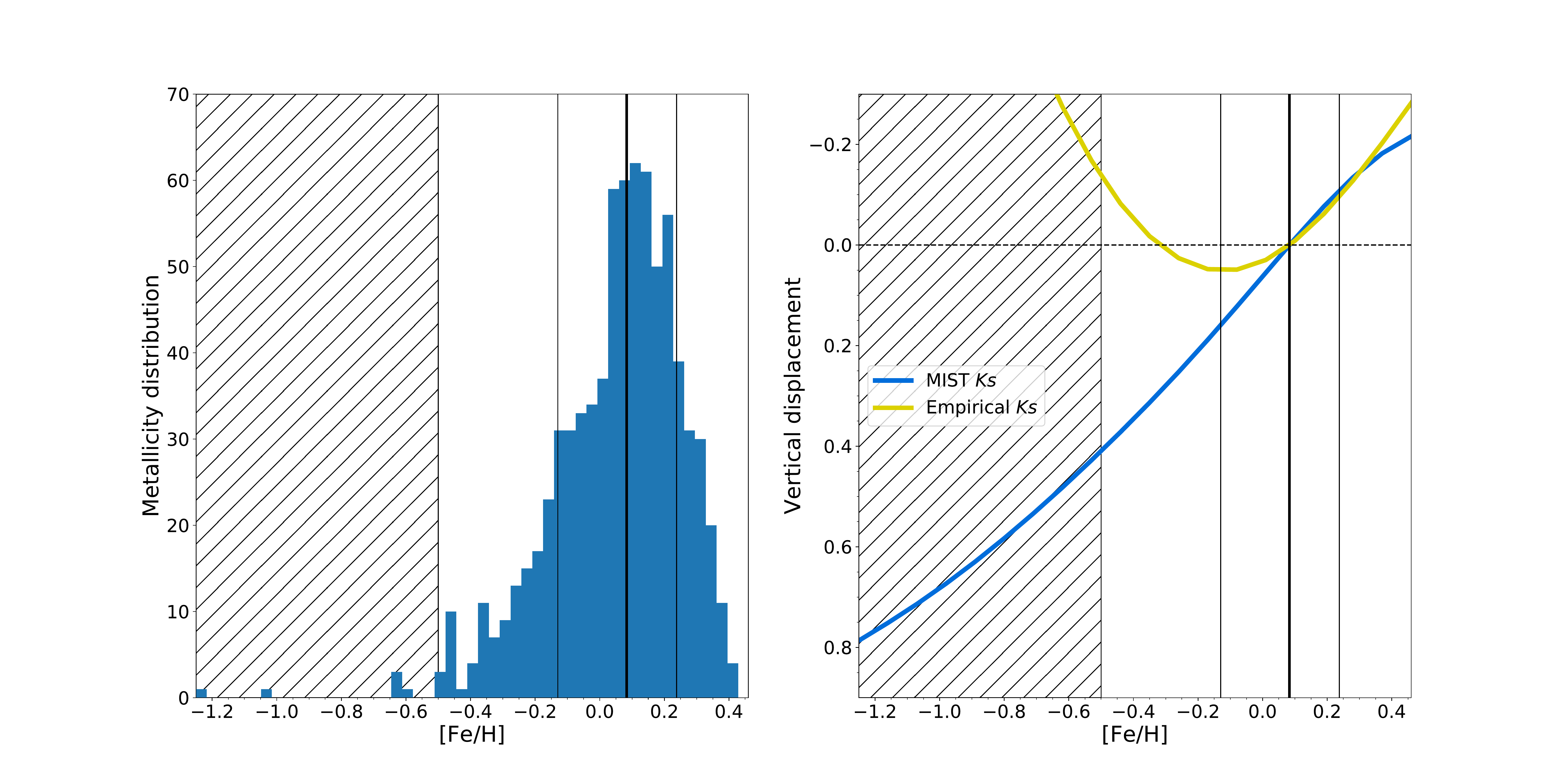}
    \caption{\emph{Left:} Metallicity distribution of the \Kepler{} cool 
        dwarfs in the APOGEE sample, selected as having 
        \(\Teff < 5250 \textrm{ K }\) and a luminosity less than 1.3 mag above 
        the single-star sequence. The 
        thick black line denotes the median metallicity while the thin black 
        lines denote the 1-\(\sigma\) confidence intervals. The hatched region
        to the left of \feh{}=-0.5 denotes the metallicity beneath which the 
        empirical correction in \cref{fig:met_trend} is poorly constrained due 
        to too few metal-poor stars. \emph{Right:} The vertical displacement 
        caused by assuming a single metallicity isochrone.  The blue curve
        represents the difference between the derived \MK{} assuming the 
        median field metallicity and that for the true metallicity. The 
        orange curve includes our empirical shape correction for metallicity, 
        which removes most of the predicted width. The hatched region
        to the left of \feh{}=-0.5 denotes the metallicity beneath which the 
        empirical correction is poorly constrained due to too few metal-poor 
        stars.}\label{fig:metallicity}
\end{figure*}

Because we will mostly be concerned with the cool dwarfs, we check the
metallicity distribution of dwarfs cooler than 5250 K, which is the 
temperature range of our sample. The metallicity distribution for that
sample is shown in the left panel of \cref{fig:metallicity}. The median 
metallicity is 0.08, consistent with that observed for planet hosts by 
the California \Kepler{} Survey \citep{Petigura17}.  

We check our assumption that the metallicity distribution for the full
\Kepler{} field is similar to the the spectroscopic catalog by comparing to
results from the California \Kepler{} Survey (CKS). The CKS observed a large
(1305) sample of \Kepler{} planet hosts, including a magnitude-limited sample
with \(Kp < 14.2\)~mag, which is substantially deeper than the APOGEE survey
\citep{Petigura17}. In order to test if the metallicity of a shallow survey, 
such as APOGEE, differs from the metallicity of a deeper survey, such as the 
CKS, we compare the metallicity distribution of the two surveys. As noted before, 
in Section~\ref{sec:speccat}, planet host surveys are likely to be biased toward 
higher metallicities. When applying the same temperature and evolutionary state
cuts to the CKS, we find that the median metallicity is 0.05 dex. A 2-sample
Anderson-Darling test finds that the two metallicity distributions are
discrepant by 2-\(\sigma\).

While the metallicity distributions of the two surveys are significantly 
different, the magnitude of the difference is small, even accounting for the 
fact that planet hosts should be more metal-rich than the overall sample. We 
therefore treat the shallow and deep samples as having the same metallicity.
The small difference in mean metallicity will be corrected for by subtracting off 
a zero-point offset, which will be done in the following sections. As a 
result, we adopt the \(\feh{} = 0.08\) isochrone as the base isochrone for the 
photometric sample. 

\subsection{Correcting Metallicity Trends}
\label{sec:speccor}

Given a large, spectroscopically characterized sample with a well-populated 
main sequence, we can empirically determine how the main sequence behaves as a
function of stellar parameters such as metallicity.  Even modest differences 
between the theoretically predicted positions of isochrones and the data can 
impose structure on the detrended main sequence. Residual trends are not 
surprising because the NIR behavior of MIST isochrones have not been 
comprehensively tested at a wide range of metallicities \citep{Choi16}.  In 
order to achieve a truly detrended main sequence, we include both metallicity 
and temperature-dependent terms.

\begin{figure}[htb]
    \centering
    \epsscale{0.9}
    \plotone{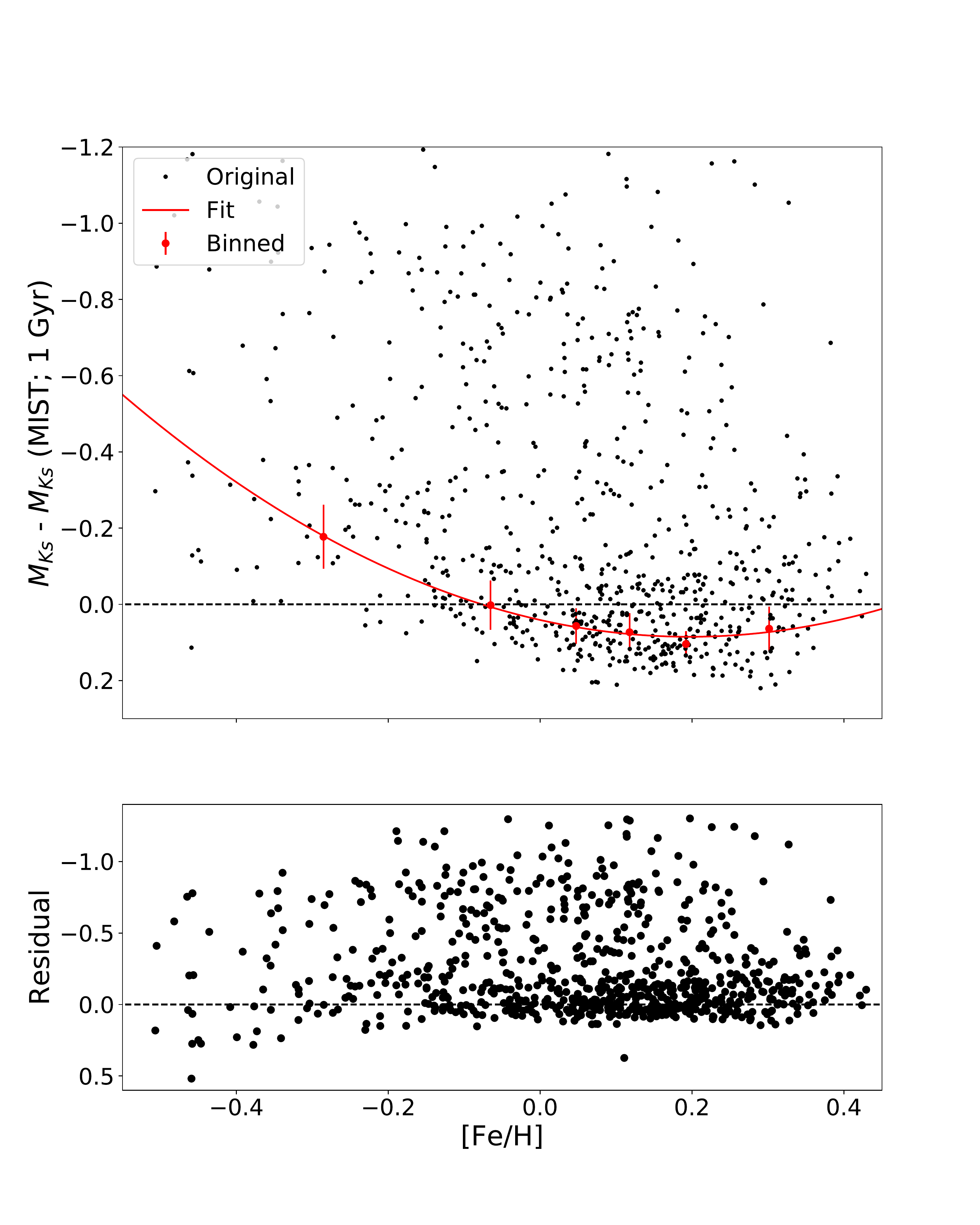}
    \caption{\emph{Top:} Residual metallicity trend in the MIST 
        metallicity-corrected vertical displacement. A quadratic trend remains over 
        metallicity. The trend is characterized by fitting the sample to 
        5 bins with equal numbers of points.  The 25th percentile of points 
        denote the y-value of the bin while the mean temperature within each
        bins denotes the x-value. The best-fit relation is \(\Delta
            \MK{} = -1.14\feh{}^2 + 0.448 \feh{} + 0.0411\). Red points mark 
            the bins used for the fit.  The median absolute deviation of the 
            50\% of the sample showing the
        least luminosity excess within each bin is shown as an error bar. Three 
        very low-metallicity points are not shown.  \emph{Bottom:} Residuals 
        after correcting for the quadratic trend.}\label{fig:met_trend}
\end{figure}

We find and correct for a residual quadratic trend in the vertical 
displacement of the main sequence over metallicity, shown in 
\cref{fig:met_trend} for the spectroscopic sample.  We fit the trend to 5 
equally populated bins. We denote 
the metallicity of the bin as the mean metallicity of all stars in that bin, 
and the vertical displacement as the 25th percentile of all stars in that bin. 
For the purpose of selecting the optimal binning style, we quantified the 
goodness-of-fit using the median absolute deviation of the lowest
50\% of the vertical displacements. The correction was mostly robust to the choice 
of binning; only the sparsely populated low-metallicity regime 
(\(\feh{} < -0.5\)) was sensitive to the details of binning. Because there are 
only eight objects in this metallicity range, our conclusions are not 
sensitive to their treatment.

\begin{figure*}[htb]
    \centering
    \plottwo{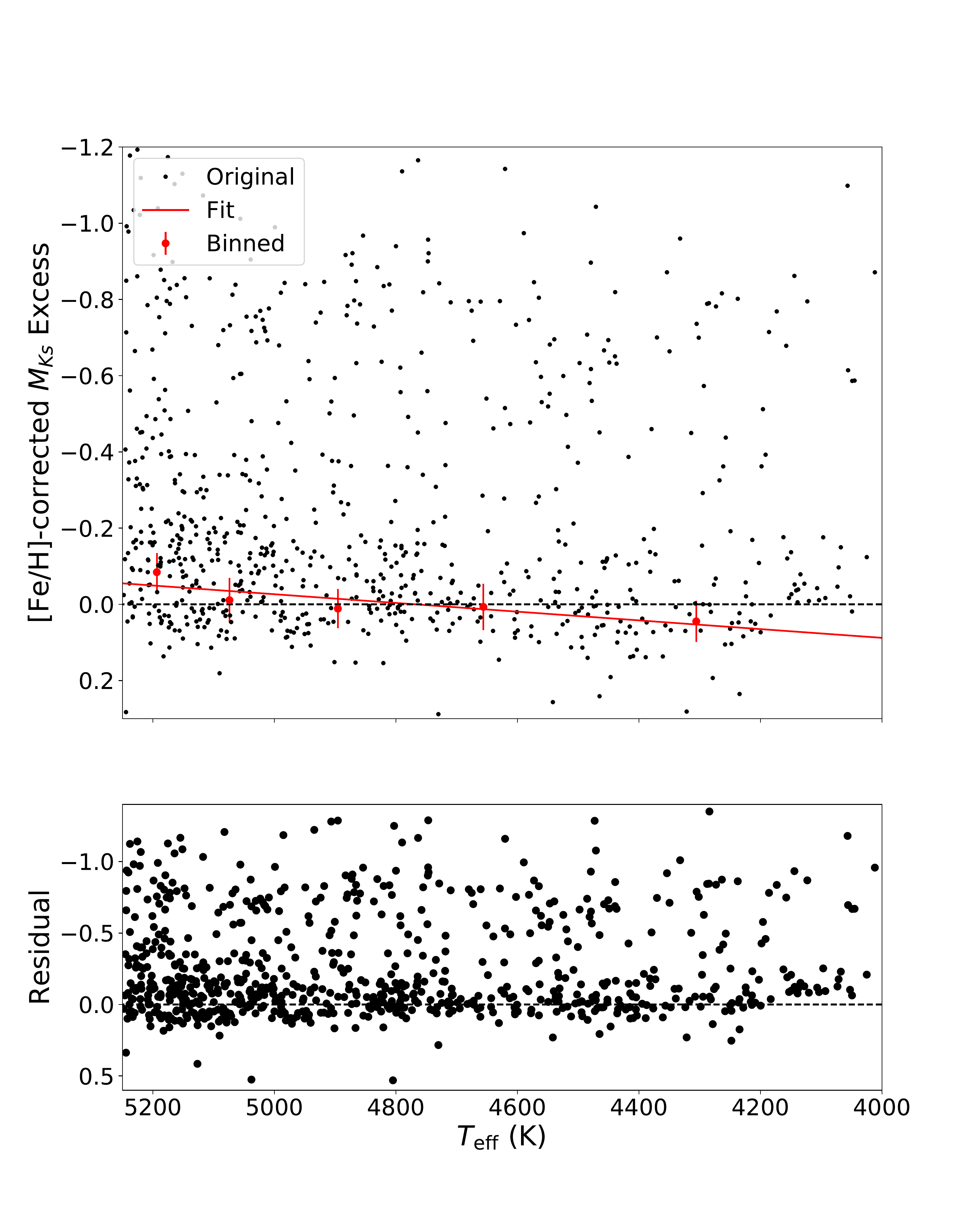}{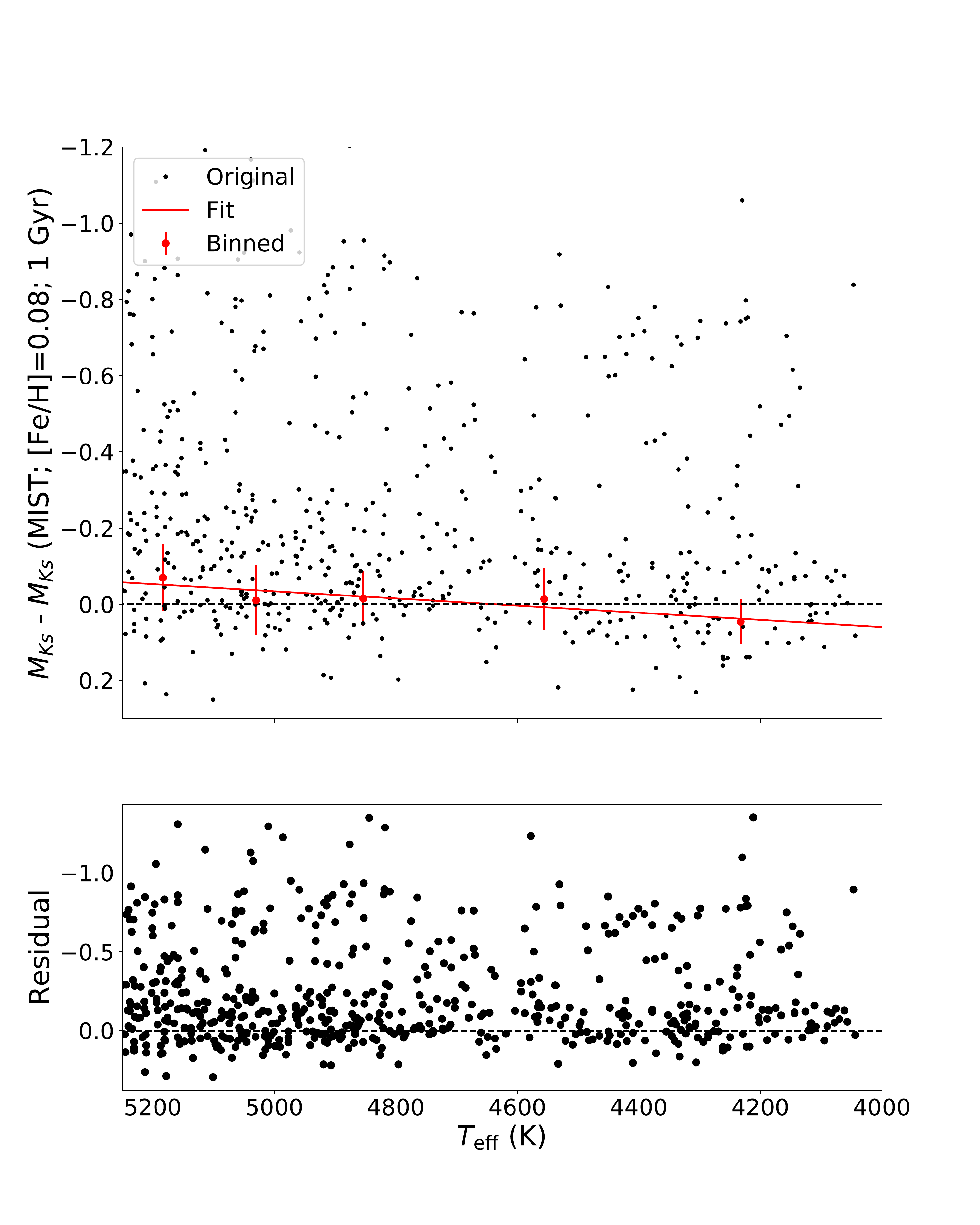}
    \caption{\emph{Top Left}: Vertical displacement above a 1 Gyr,
        metallicity-adjusted MIST isochrone, empirically corrected for
        metallicity trends, for the cool sample with APOGEE temperatures. 
        A slight linear trend in \Teff{} was measured using five bins (shown in
        red). The best-fit relation is \(\Delta \MK{} = -0.000114
        \Teff{} + 0.546\) The median absolute deviation of the 50\% of the 
        sample showing the least luminosity excess within each bin is shown as 
        an error bar. \emph{Bottom Left:} The residuals after subtracting the 
        linear trend. Not shown are the two most metal-poor objects in the
    sample, which have substantially large 
residuals. {Top Right, Bottom Right:} Same as left except using
\citet{Pinsonneault12} temperatures, and subtracting off a constant
\feh{}=0.08 isochrone for all objects instead of a metallicity-adjusted
isochrone. The best-fit relation is \(\Delta \MK{} = -0.0000937 \Teff{}
+ 0.434\)}\label{fig:apogee_teff_trend}
\end{figure*}

Combining the two changes: subtracting only a single-metallicity isochrone and
using the photometric temperatures yields the points in the right panel of 
\cref{fig:apogee_teff_trend}.
This sample still has a trend with temperature, that we remove using a linear
fit, similar to that done for the spectroscopic sample.

\subsection{Temperature Corrections}

\subsubsection{Spectroscopic Sample}

After correcting the trend in metallicity, we correct a residual trend in
\Teff{}, shown in \cref{fig:apogee_teff_trend} using a similar procedure. We
take five equally-populated bins, and define the bin temperature and 
displacement using the mean within the bin  and 25th percentile displacement
within the bin.  This trend is slight and can be subtracted using a linear 
fit. After performing these two corrections, we consider the main sequence to be 
flattened to its measured natural width, as seen in the bottom panel of 
\cref{fig:apogee_teff_trend}. The persistence of the binary sequence in the 
residuals indicates that our corrections are meaningful.  

\subsubsection{Photometric Sample}

The photometric sample in \citet{McQuillan14} differs from the spectroscopic
sample in two main ways: the lack of metallicity diagnostics, and the necessity
of using photometric instead of spectroscopic \Teff{}. Both of these
differences will increase the uncertainty of the inferred single-star 
\(Ks\)-band luminosity, and we measure their impact in the following analysis.

\begin{figure}[htb]
    \centering
    \plotone{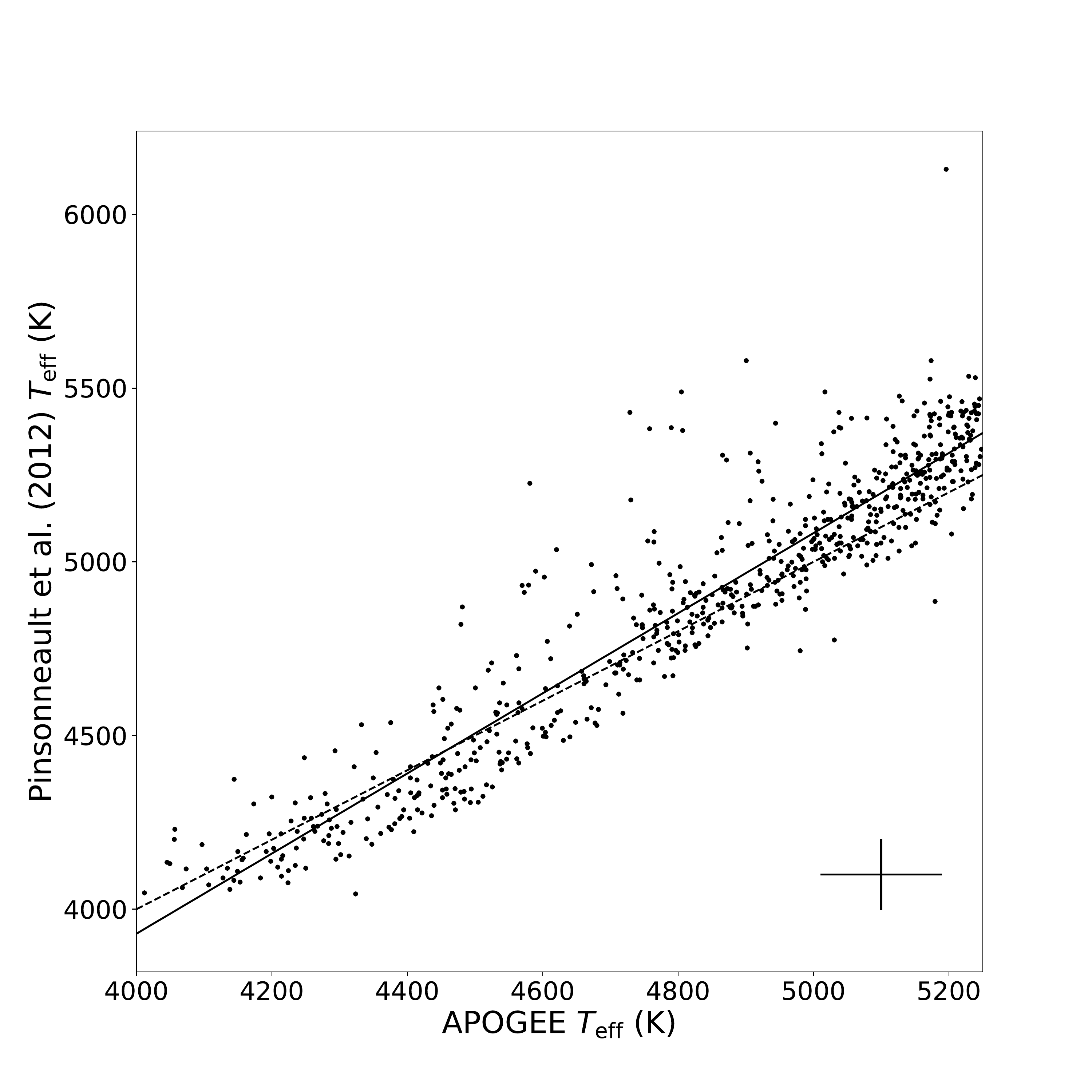}
    \caption{Comparison between the APOGEE spectroscopic \Teff{} and the
        \citet{Pinsonneault12} \Teff{} for objects with \(4000 \textrm{ K } \le
    \textrm{ APOGEE \Teff{} } \le 5250 \textrm{ K }\). A linear fit for the
    temperature is shown as the solid line. The best-fit model is
    \(T_{\textrm{eff,phot}} = 1.153 T_{\textrm{eff,spec}} - 682.6\). The 
    one-to-one line is shown as a dashed line. The scatter between the two 
    datasets is 135 K. A representative error for both datasets is shown in 
the bottom-right corner.}\label{fig:teffdiff}
\end{figure}

We estimate the impact of the photometric temperatures by directly 
comparing the spectroscopic and photometric temperatures for the set of 
overlapping points shown in \cref{fig:teffdiff}. Overall, the two temperature scales are reasonably
correlated. There is a temperature scatter of 135 K between the two scales.
Taking into account the 100 K uncertainty in the APOGEE \Teff{} scale, by
assuming the two uncertainties add in quadrature, this
implies a 90K uncertainty in the \citet{Pinsonneault12} temperatures, which
agrees with the uncertainties expected from the photometric method. 

\subsection{The Impact of Metallicity on the Main Sequence Width}
\label{sec:mswidth}

\begin{figure*}[htb]
    \centering
    \plotone{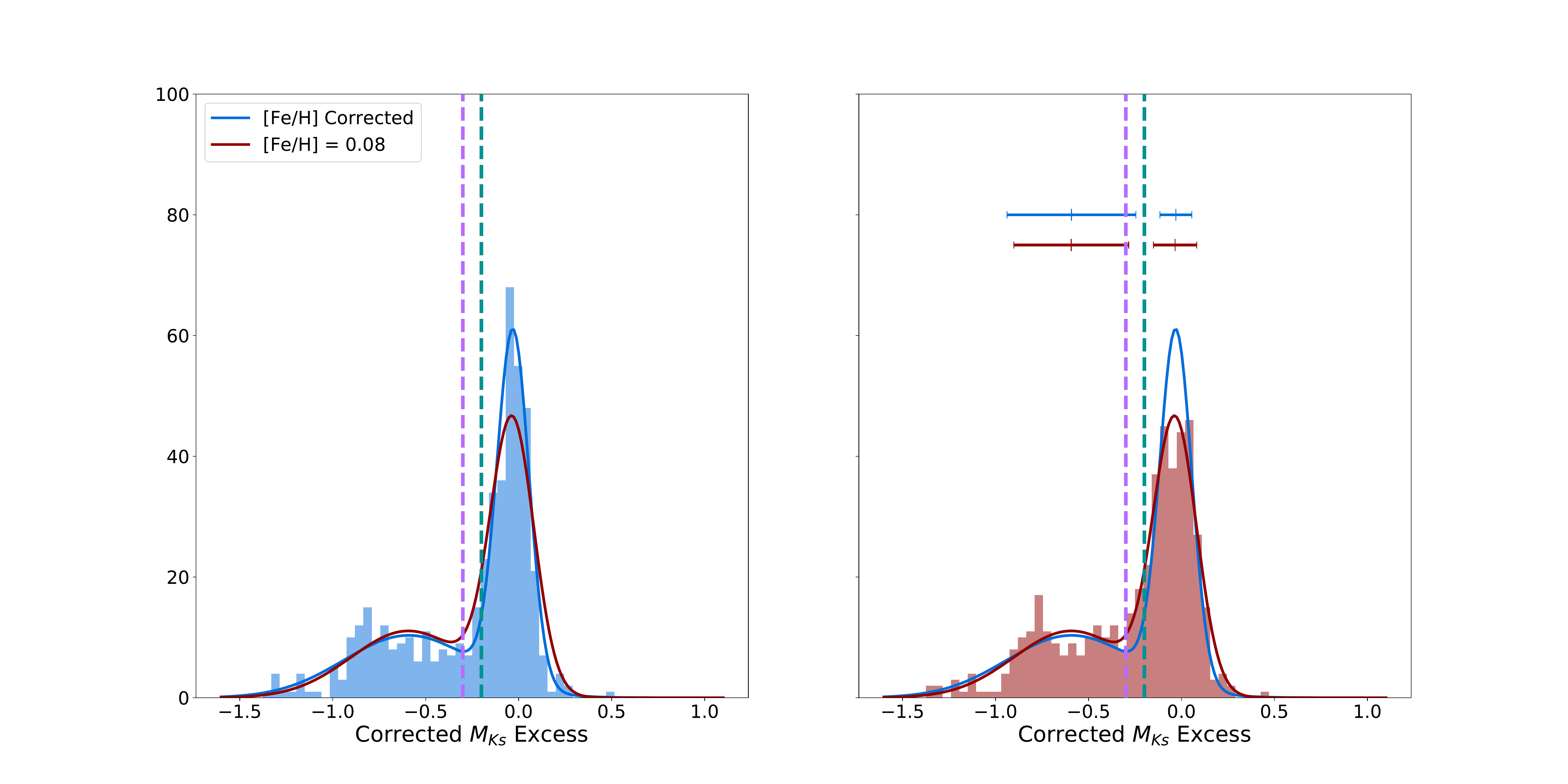}
    \caption{\emph{Left:} Histogram showing the distribution of vertical 
        displacements with isochrones corresponding to APOGEE metallicities.
        The blue curve denotes the best-fit double-Gaussian model to the
        distribution of vertical displacements. The red curve is a fit to the
        best-fit double-Gaussian model assuming a single metallicity (see right
        panel). Conservative and inclusive photometric binary thresholds at 
        \(\Delta \MK{} < -0.3\) mag and \(\Delta \MK{} < -0.2\) mag are shown as 
        violet and green dashed lines, respectively. The best-fit dispersion of 
        the single-star Gaussian is 0.086 mag. \emph{Right:} Histogram showing 
        the distribution of vertical displacements with a \(\feh{} = 0.08\) dex
        isochrone. The red curve denotes the best-fit double-Gaussian model to
        the distribution of vertical displacements. The blue curve is a fit to
        the best-fit double-Gaussian model for isochrones adjusted for the
        APOGEE metallicity (see left panel). Conservative and inclusive
        photometric binary thresholds are shown as violet and green dashed 
        lines, respectively. The best-fit dispersion of the single-star 
        Gaussian is 0.117 mag. The blue points at the top of the plot show the
    mean and standard deviation of the single and binary Gaussian fits for the
sample with metallicity information. The red points show the mean and standard
deviation of the single and binary Gaussian fits for the sample without
metallicity information.}\label{fig:histcompare}
\end{figure*}

Now that we have corrected vertical displacements over the 4000--5250 K
temperature range, we can now compare the main-sequence width for the two
datasets. Histograms for the spectroscopic and photometric samples over 
vertical displacement are shown in \cref{fig:histcompare}. We try to 
characterize the width of the two samples by fitting a double Gaussian to each 
histogram, and taking the width of the Gaussian centered on the main
sequence as the uncertainty in the vertical displacement. Our fits indicate that 
the main-sequence width for the spectroscopic sample is 0.086 mag while the width 
for the photometric sample is 0.117 mag.

\begin{figure*}[htb]
    \centering
    \plotone{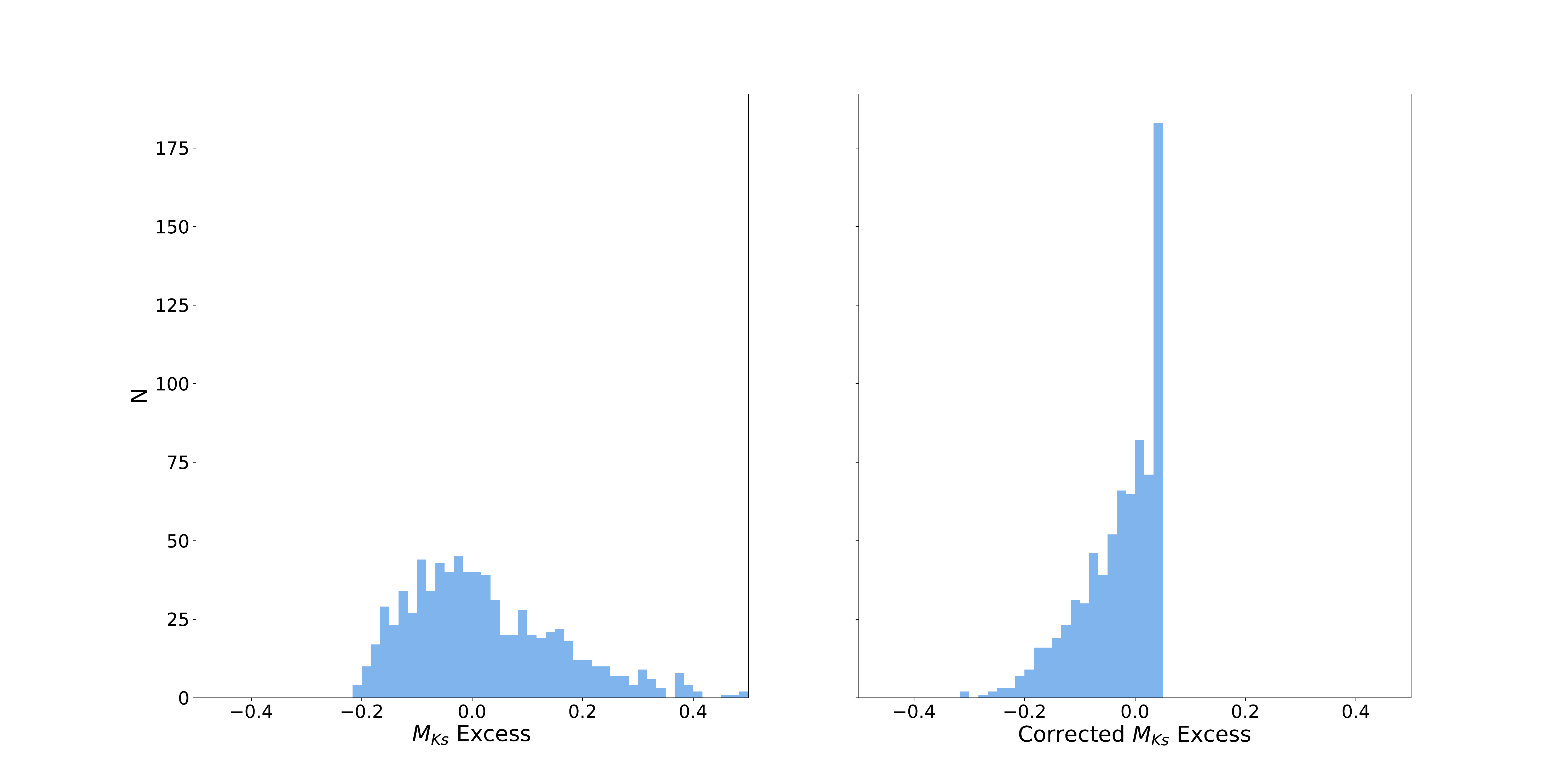}
    \caption{\emph{Left:} The distribution of vertical displacements imposed
    by the metallicity distribution of the APOGEE sample using the raw MIST
    isochrones. All stars are assumed to have a known temperature of 5000 K. 
    Not shown are two very metal-poor stars which are off the plot to the 
    right. \emph{Right:} Same as the left, except the MIST \(Ks\)-band 
    magnitudes are corrected by the empirical correction in
\cref{fig:met_trend} and the two most metal-poor stars are off the figure to
the left.}\label{fig:met_spread}
\end{figure*}

This modest increase in the width for the photometric sample is surprising,
especially given the strong dependence of \(Ks\)-band magnitude on metallicity 
predicted by the MIST isochrones shown in the right panel of 
\cref{fig:metallicity}. We believe that the reduced width is real, and that 
the MIST bolometric corrections overestimate the dependence of \(Ks\)-band 
luminosity on metallicity. We demonstrate the metallicity dependence of 
the isochrones by measuring the 
main-sequence width for an artificial population of stars which 
differ only by metallicity. We assume all stars have a temperature of 5000 K, 
an age of 1 Gyr, and have metallicities drawn from the APOGEE cool dwarf 
metallicity distribution (shown in the left panel of \cref{fig:metallicity}).

We display two realizations of this population in \cref{fig:met_spread}: one
using \(Ks\)-band magnitudes predicted by MIST (left), and another applying the
metallicity-dependent correction in \cref{fig:met_trend} (right).
The main-sequence width of the two populations are 0.14 mag and 0.10 mag, 
respectively, indicating that the corrected distribution has a smaller width. 
As is evident from the corrected sample on the right side of 
\cref{fig:met_spread}, the reduction in the width of the main 
sequence is driven by a cutoff at large, positive vertical displacements.
The origin of this cutoff comes from the metallicity-dependence of the empirical 
correction in \cref{fig:metallicity}; metal-poor stars are forced to smaller 
vertical displacements compared to the MIST predictions. We note that the MIST 
models predict that the metallicity dependence \(Ks\)-band is similar to the
\(V\)-band, which indicates that the predicted width is a feature of the 
luminosity dependence of the main sequence locus, not a peculiar feature of 
the \(Ks\)-band bolometric corrections.  

\subsection{Characterizing the EB Distribution}
\label{sec:ebmodel}

\begin{figure}[htb]
    \centering
    \plotone{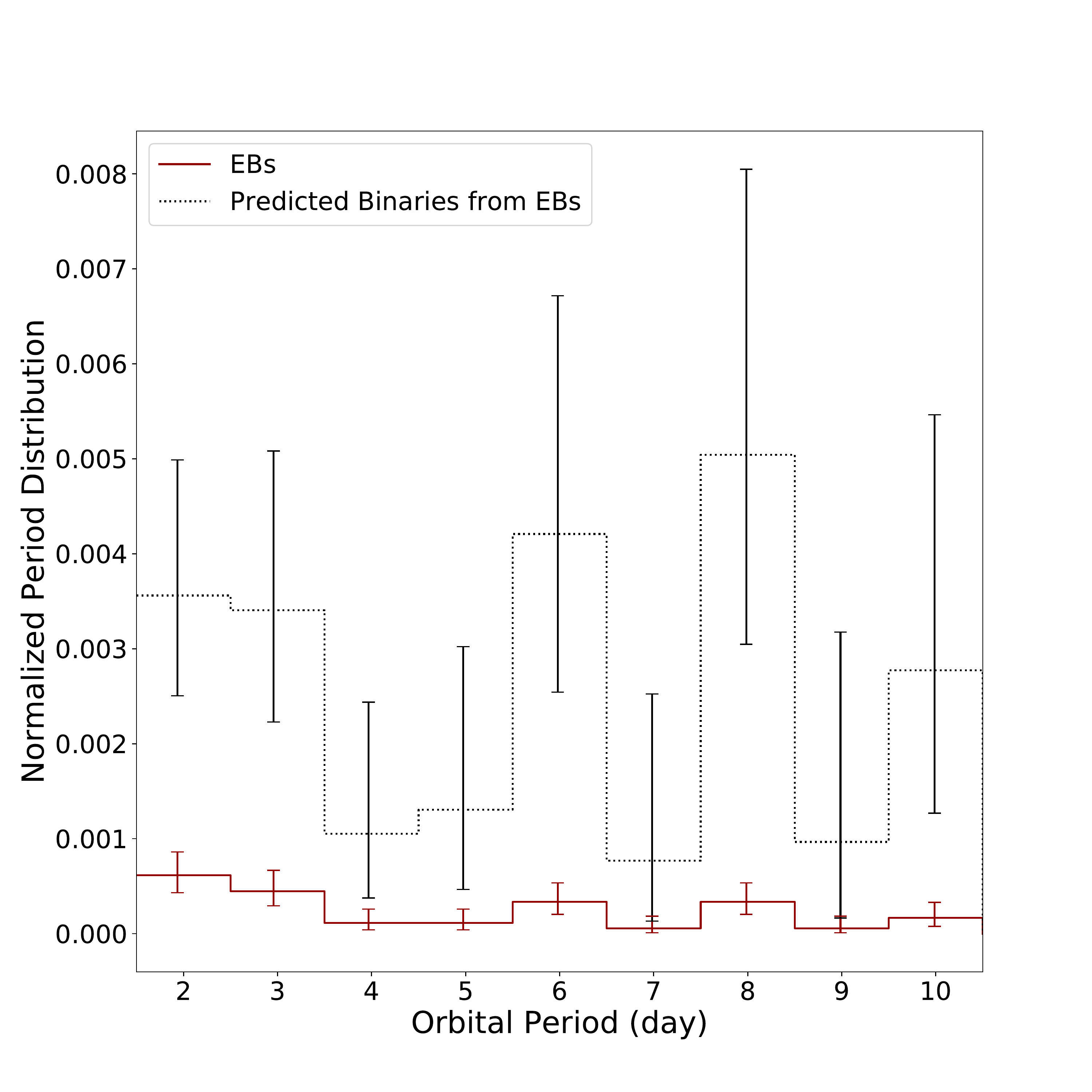}
    \caption{The observed period distribution of all eclipsing binaries with
    \citet{Pinsonneault12} \Teff{} < 5250 K and vertical displacement less
than 1.3 mag is shown in red. The period distribution of the full binary
distribution inferred from the eclipsing binaries is shown in black. The
normalization factor is the total number of Kepler cool dwarfs. Error bars
represent 1-\(\sigma\) Poisson confidence intervals.}\label{fig:ebdist}
\end{figure}

\Kepler{}'s short 30 minute cadence allows it to detect essentially all 
eclipsing main-sequence binaries with periods less than 10 days \citep{Kirk16}. 
Even the shortest-period systems become detectable through ellipsoidal variations 
rather than through eclipses. As a result of \Kepler{}'s completeness at these 
short periods, the distribution of eclipsing
binaries is solely determined by geometry, with an eclipse probability for circular 
orbits given as \((R_1 + R_2)/a\), where \(R_1\) and \(R_2\) are the primary 
and secondary radii, and \(a\) is the semimajor axis of the orbit. This simple
expression means that using the period distribution of the eclipsing binaries, 
as well as estimates for the masses and radii of the component stars, one can
predict the expected period distribution of the full binary sample using just 
the eclipse probability. We use the expression assuming circular orbits, which 
is justified because short-period, low-mass systems are overwhelmingly observed 
to be circularized \citep{Raghavan10,VanEylen16}. 

We infer the primary radius by mapping temperature to radius using the MIST
isochrones. Depending on whether the analysis is done in the context of the
spectroscopic sample or the photometric sample, the appropriate temperature
is used.

To simplify the calculation, we assume that the secondaries are half as massive 
as the primaries, which is generally consistent with a flat mass-ratio 
distribution \citep{Raghavan10}. More massive secondaries would reduce the 
predicted number of non-eclipsing systems. 

While the semimajor axis isn't directly observable, we infer it from the 
orbital period using \Kepler{}'s third law, which requires masses for the two
components. We infer masses using a method similar to the radii above: the
primary mass is mapped from \Teff{}, while the secondary mass is assumed to be
half of the primary's mass.

Given these assumptions, an estimate of the orbital period distribution 
for all short-period \Kepler{} binaries on the unevolved lower main sequence is 
shown in \cref{fig:ebdist}. The eclipsing binaries range from making up almost 
15\% of the total population of binaries in the shortest-period bin in 
\cref{fig:ebdist} to 6\% in the longest-period bin, hence they are not a 
negligible segment of the population. \citet{McQuillan14} excluded eclipsing 
binaries from their analysis of rotation in the \Kepler{} field due to concerns 
that the eclipse signal would dominate the period. In order to obtain an 
unbiased view of the binary population at short periods, we calculate vertical 
displacements for the \Kepler{} eclipsing binaries as well and include them in 
our statistical analysis in Section~\ref{sec:results}. This treatment assumes that 
orbital period is identical to the rotation period, which should certainly be 
true for synchronized systems; we address the question of synchronicity 
directly in Section~\ref{sec:synch}. By combining the bins in
\cref{fig:ebdist}, the eclipsing binary population predicts that 
\(2.1\% \pm 0.4\%\) of the cool \Kepler{} targets should be binaries with
periods between 1.5--10 days.

\subsection{Binarity and Temperature Estimation}

\begin{figure*}[htb]
    \centering
    \plottwo{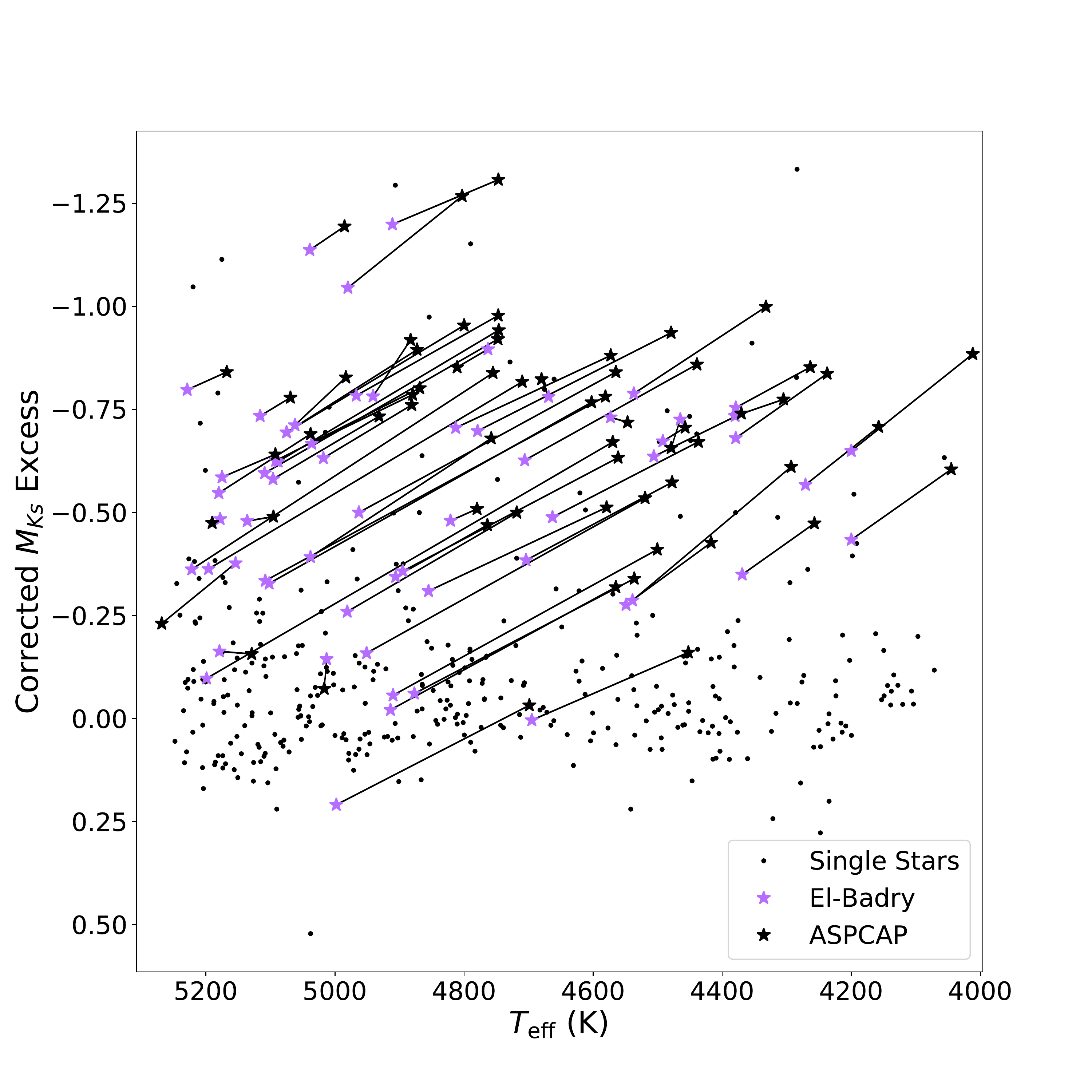}{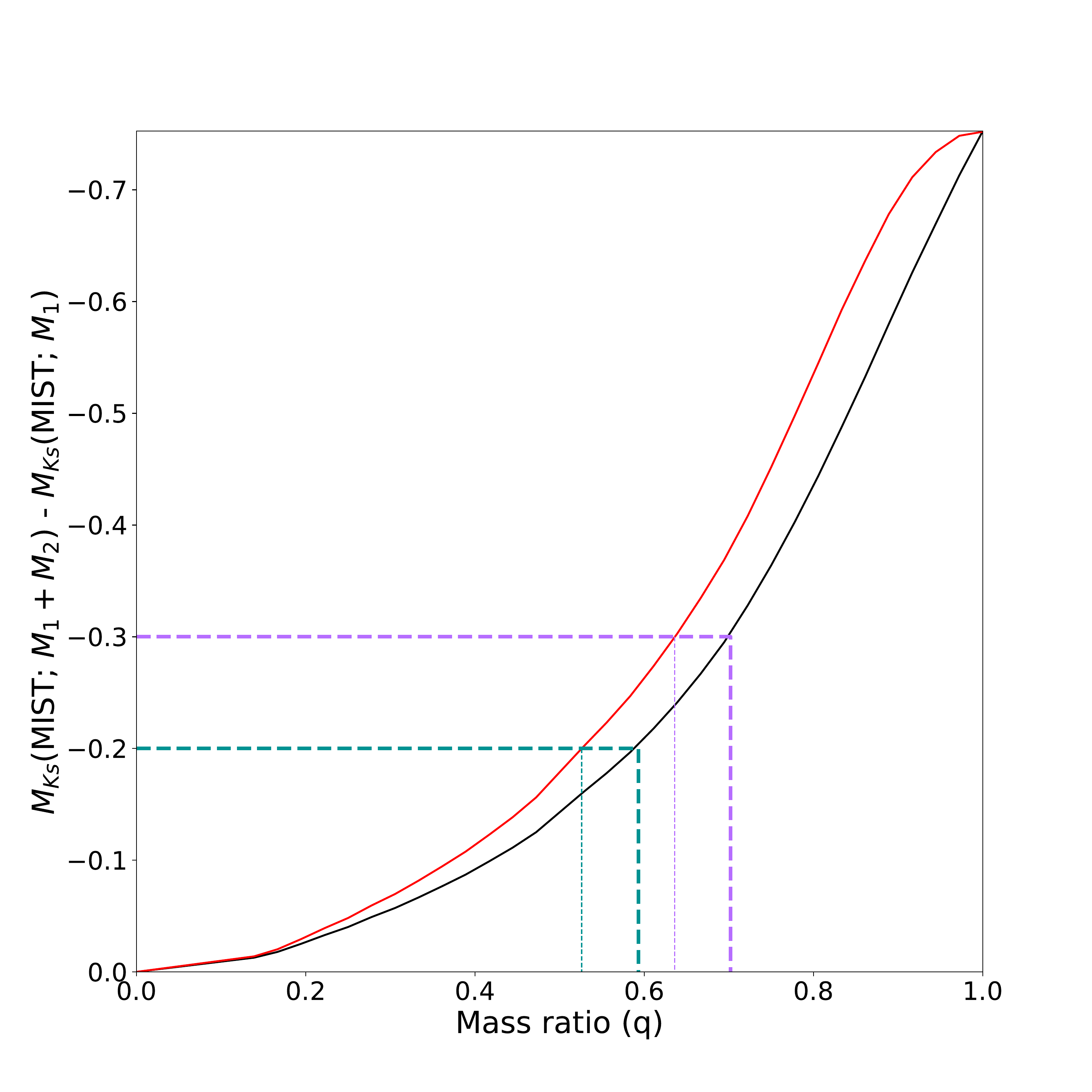}
    \caption{\emph{Left:} The sample of APOGEE targets analyzed by 
        \citet{ElBadry18b} for
    spectroscopic signs of a companion. Black points indicate the APOGEE DR14 
values for the targets. Violet points indicate the revised temperatures, and the
\(Ks\)-band luminosity excess derived from the revised temperature and 
metallicity in \citet{ElBadry18b}. \emph{Right:} Vertical displacement of a 
binary with a 0.7 M\(_\sun\), solar metallicity primary as a function of mass 
ratio using MIST models (black line). The violet and green dashed lines denote 
the mass-ratio at which the vertical displacement exceeds -0.3 and -0.2  mag, 
which are mass-ratios of 0.69 and 0.58. The red line denotes the same vertical 
displacement, but taking into account distortions in the photometric \Teff{} 
caused by binarity. The mass-ratios corresponding to the conservative and 
inclusive thresholds are 0.64 and 0.53. }
    \label{fig:binary_teff}
\end{figure*}

While we generally assume that objects showing large vertical displacement are
binaries, our sample includes an abundance of systems with vertical
displacements greater than 0.75 mag, some greater than even 1.3 mag. One
obvious explanation for these systems is that they are multiple systems 
where each component contributes substantially to the luminosity. One other 
explanation is that the secondary biases the observed temperature to be 
cooler than the actual primary temperature causing the single-star luminosity 
to be underestimated. This phenomenon is expected to occur for both 
photometric and spectroscopic temperatures \citep{Pinsonneault12,ElBadry18a}. 

We test the effect of binarity on the estimated spectroscopic temperature by 
using APOGEE DR13 targets whose spectra were fit for multiple components
\citep{ElBadry18b}. In cases where a two-component fit was favored,
temperatures were published for both the primary and secondary. With \Gaia{} DR2 
parallaxes, we can test how often two photometric components were successfully
detected. All the DR13 objects analyzed by \citet{ElBadry18b} which overlap
with the cool dwarf rotation sample are shown in
\cref{fig:binary_teff} with both the ASPCAP and decomposed primary 
temperatures.

The overall behavior in the \citet{ElBadry18b} sample shows that the effect of 
binarity on spectroscopic temperatures is substantial. For many of the 
intermediate mass-ratio objects, there is a temperature discrepancy of several 
hundred Kelvin, leading an overestimation of the \(Ks\)-band vertical 
displacement by up to 0.3 mag. The difference in temperature is less severe for 
targets with high luminosity excesses, which is to be expected because the two 
stellar components would have similar spectra. However, the effect is still 
not negligible. For the purpose of flagging photometric binaries, this effect
is beneficial, as it increases the contrast of lower mass-ratio systems
against the single stars. To characterize the luminosity-ratio of the 
components, however, the offset of temperatures would need to be taken into 
account. 

While the two-component fit produces results that overall seem reasonable, many
of its decompositions were not compatible with the vertical displacement. Many
of the photometric binaries were not flagged as requiring two-component fits.
Additionally, two of the targets showing essentially no vertical displacement
were flagged as having substantial modifications to the temperature from a
companion. Nonetheless, we consider the results of the analysis by 
\citet{ElBadry18b} to impressive given that they did not make use of \Gaia{} 
parallaxes. Future analyses may want to make use of the vertical displacement 
when flagging spectra for multiple-component fits.

We also investigate the impact of binarity on the vertical displacement using
photometric temperatures. Unlike \citet{ElBadry18b}, who directly fit 
two-component models to the full spectra, we instead are interested in how 
companions generically affect the vertical displacement through temperature. We
explore by building a toy model of a binary using MIST isochrones, and perturbing 
the observed temperature using a color-\Teff{} relation. Because the 
\citet{Pinsonneault12} temperatures were derived using \textit{griz} colors, we 
opt to use the \((g-i)\)-\Teff{} relationship to probe the impact of binarity on 
the temperature. Redder colors are more sensitive to low-mass binary companions, 
so we choose the \( (g-i) \) color as the reddest color that maintains high 
photometric precision. The effect of binaries should be less pronounced for
bluer color-temperature relations.

To isolate the effect of temperature offsets on vertical displacement, we adopt a 
standard primary with a mass of 0.75 \(M_\sun\) and solar metallicity, 
and use the MIST isochrones to calculate an intrinsic \(g-i\) color. We
derive a temperature for the star by fitting a 5th-degree polynomial between
\(\tfrac{5040 K}{\Teff}\) and \((g-i)\) using the MIST synthetic photometry. We
then generate secondaries with a range of mass-ratios, and add their flux
to the primary flux, and infer a composite temperature based on the combined
flux. Lastly, the composite temperature is then used to calculate the reference
K-band magnitude as opposed to the true primary temperature. The relationship
between vertical displacement and mass-ratio without taking into account the
color of the secondary is shown in black in \cref{fig:binary_teff}. The relationship behaves as expected,
with a sharp increase at high mass-ratio until 0.75~mag. When the color of the 
secondary is taken into account, we get the red curve in the left panel of 
\cref{fig:binary_teff}. The impact of the binary on photometric temperatures is
substantially smaller than for the spectroscopic temperatures, only increasing
the vertical displacement by up to 0.1~mag. Our toy model also predicts that
the temperature deviations are so small compared to the slope of the main
sequence that the vertical displacement never exceeds 0.75~mag, as is observed in
the spectroscopic case.

Because the effect of binarity on photometric temperature only imparts small 
deviations for the vertical displacement, we claim that the objects which lie
substantially above the photometric binary sequence are true multiple systems,
or systems with an unusual evolutionary history, such as mass-transfer from a
companion. 

\section{Results and Discussion}
\label{sec:results}

\subsection{Vertical Displacements of Rapid Rotators}
\label{sec:elbadry}

After successfully calculating the vertical displacement of the full sample, we
now attempt to distinguish between single and binary stars. As noted in
Section~\ref{sec:mswidth}, the 1-\(\sigma\) width of the single-star sequence 
were 0.086 mag with known metallicity, and 0.117 mag when metallicity was unknown. 
To implement a uniform photometric binary cut for both the spectroscopic and
photometric samples, we will assume a measured width of the main sequence of 
0.1 mag. To prevent age effects 
from biasing the results, we only performed statistics in the 4000--5250 K 
range.

The primary quantity we use to characterize the binarity of our samples is the
photometric binary fraction, defined as the fraction of targets within a given
period range that have a vertical displacement above a certain threshold.
To show that our findings are robust to the choice of photometric
binary threshold, we report photometric binary fractions for two photometric 
binary thresholds: an inclusive 2-\(\sigma\) threshold and a conservative 
3-\(\sigma\) threshold, which turn out to be displacement cuts of 0.2 and 0.3 
mag. 

\begin{figure*}[htb]
    \centering
    \epsscale{1.1}
    \plotone{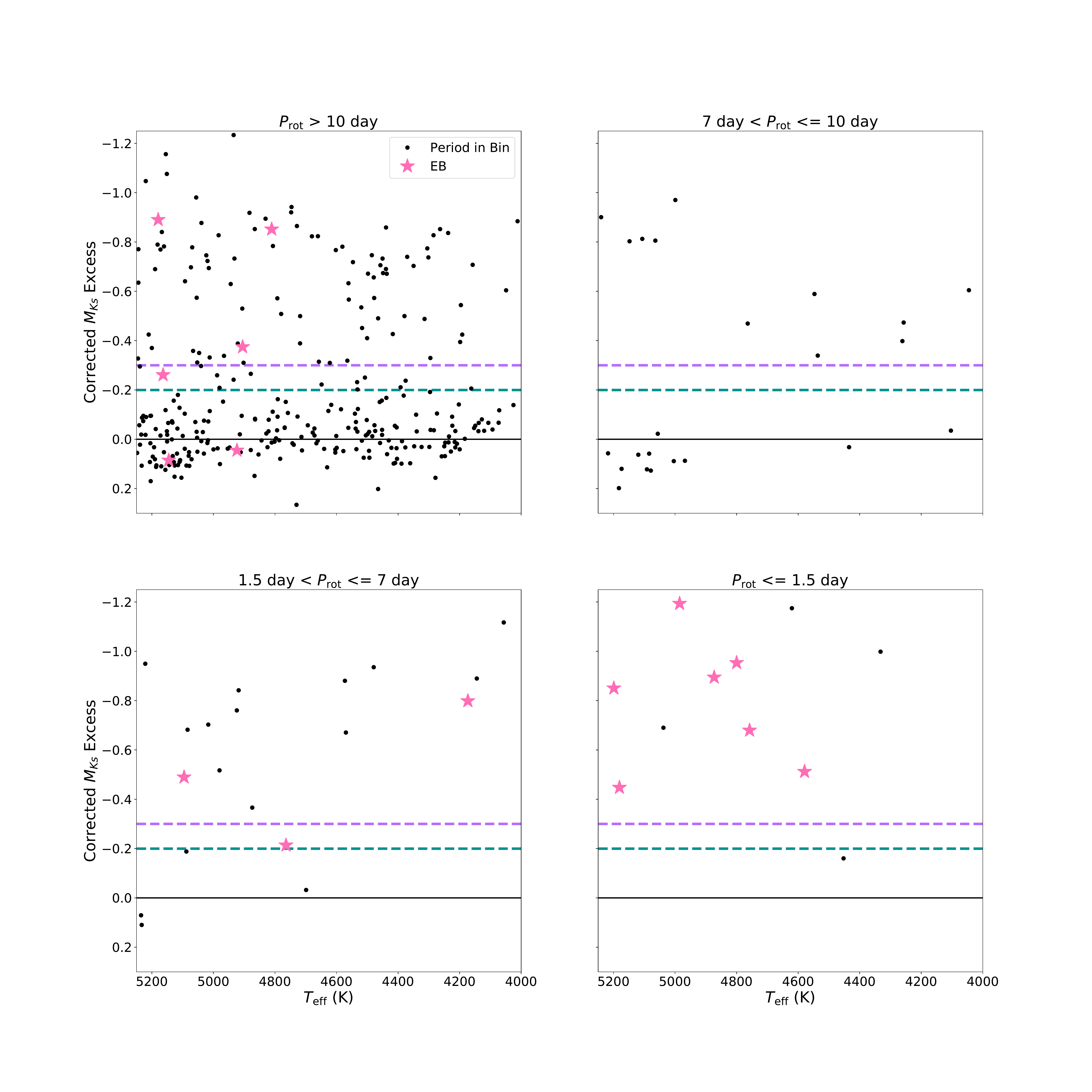}
    \caption{\emph{Top Left to Bottom Right:} Vertical displacement of cool 
        APOGEE targets with \citet{McQuillan14} periods >10 days, between 
        7--10 days, 1.5--7 days, and <1.5 days. Pink stars denote 
        eclipsing binaries with orbital periods within the same ranges. The 
        green and purple lines denote the inclusive and conservative 
        photometric binary thresholds, respectively. The temperatures are from
    APOGEE.}\label{fig:apogee_rapid_excess}
\end{figure*}

For both binarity thresholds, the well-characterized APOGEE sample 
illustrates the differences in binarity between slow and rapid rotators, as
shown in \cref{fig:apogee_rapid_excess}. For
the purpose of illustration, we divide the sample into four period ranges, with
boundaries of 1.5, 7, and 10 days. We draw a short-period limit for our analysis 
at 1.5 days because we want to eliminate potential contamination from 
ellipsoidal variables and semidetached systems \citep{VanEylen16}. We find 7 
days to mark the transition from a binary-dominated synchronized population to 
a single-star dominated population (see below). And 10 days is the 
theoretically and observationally-motivated boundary in orbital period where 
synchronization should take place \citep{Claret97,Lurie17}.

Even by eye, it is striking in \cref{fig:apogee_rapid_excess} that nearly all 
of the rapid rotators with periods between 1.5--7 days are photometric 
binaries, in contrast to the slow rotators with periods greater than 10 days, 
which have a distinct single star sequence.  The depletion of the single-star 
sequence in the short-period regime is likely due to angular momentum loss due 
to stellar winds acting over a billion years. The transitional periods between 
7--10 days have a binary fraction intermediate to the rapid and slow regimes. 
The bin with periods shorter than 1.5 day is not expected to behave like 
detached, eclipsing binaries due to potential contamination from contact 
binaries, ellipsoidal variables and blended pulsators \citep{VanEylen16}. 

\begin{figure*}[htb]
    \centering
    \epsscale{1.1}
    \plotone{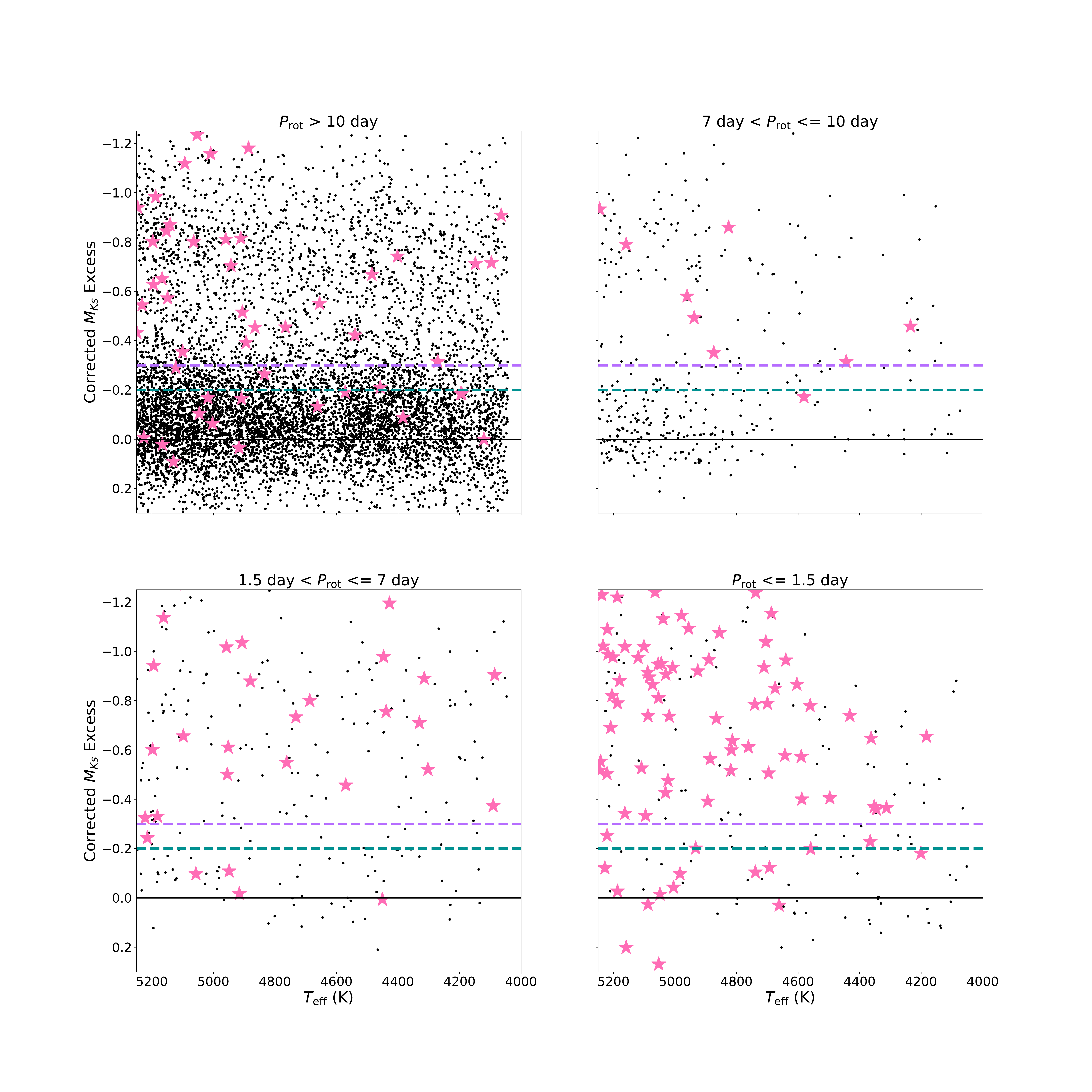}
    \caption{\emph{Top Left to Bottom Right:} Vertical displacement of all 
        cool \citet{McQuillan14} targets in period ranges >10 days, 
        7-10 days, 1.5-7 days, and <1.5 days. Pink stars 
        denote eclipsing binaries with orbital periods within the same ranges. 
        The green and purple lines denote the inclusive and conservative 
    photometric binary thresholds, respectively. The temperatures are from
\citet{Pinsonneault12}.}\label{fig:mcq_rapid_excess}
\end{figure*}

While the spectroscopic sample shows broad trends between rotation and binarity 
in wide period ranges, the substantially larger photometric sample is needed to 
effectively characterize the rapid rotators. The vertical displacement of the
photometric sample in the same period ranges as before is shown in 
\cref{fig:mcq_rapid_excess}.  With a substantially larger number of objects, the 
absence of the single-star sequence in the panel of rapid rotators is starker. 

\begin{figure*}[htb]
    \centering
    \epsscale{1.1}
    \plotone{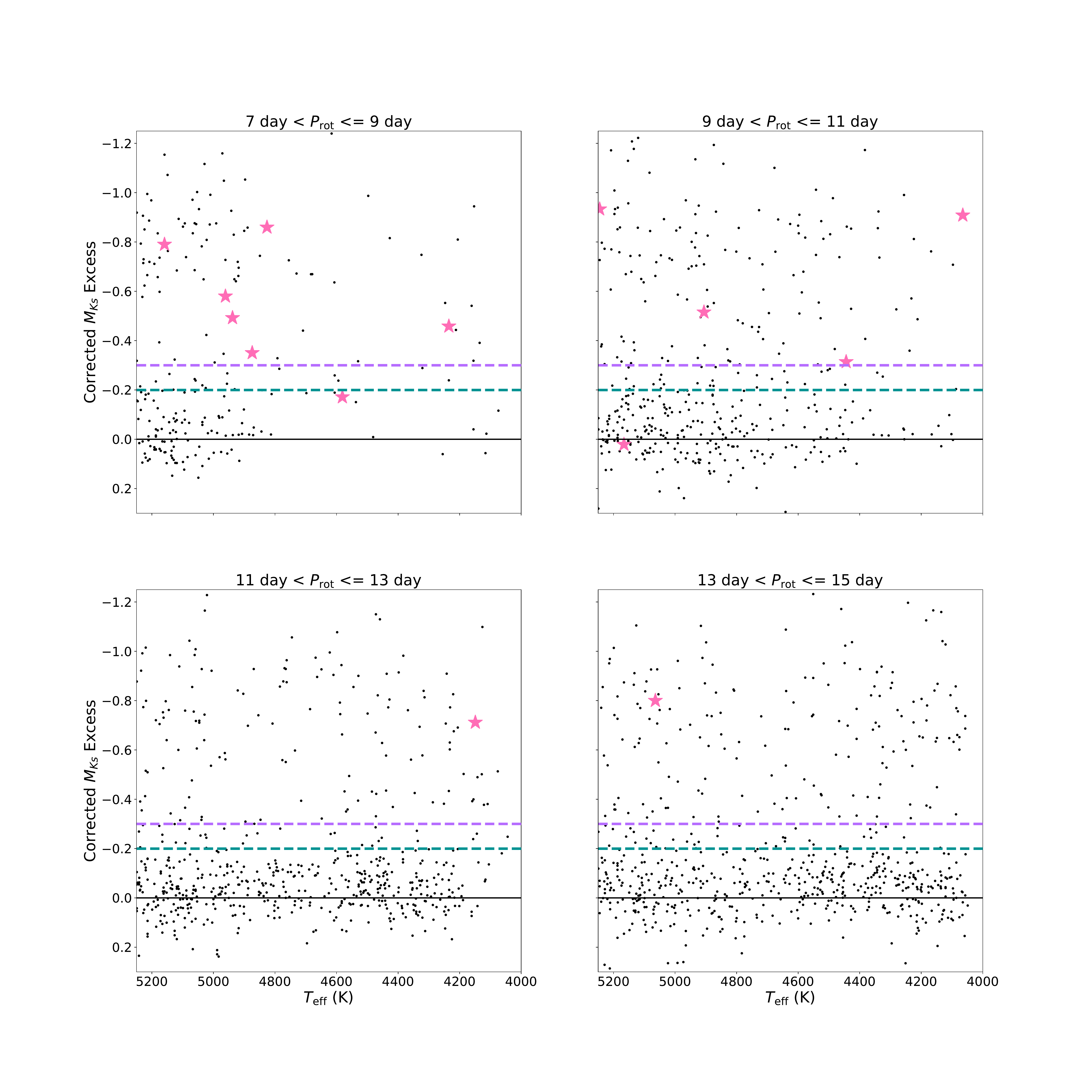}
    \caption{Same as \cref{fig:mcq_rapid_excess} except with the period ranges
    7--9 days, 9--11 days, 11--13 days, and 13--15 days.}
    \label{fig:mcquillan_transition}
\end{figure*}

The behavior of the single-star sequence is shown in
\cref{fig:mcquillan_transition}, where we show a narrower range of
periods in the transition region. The single-star sequence appears in the
hottest stars first in the panel with the most rapidly rotating systems. Cooler
stars appear on the single-star sequence as longer periods. The trend
that hotter stars in the field rotate more quickly than the cooler stars 
is consistent with the overall trend in temperature seen in 
\citet{McQuillan14}.

\begin{figure*}[htb]
    \centering
    \plotone{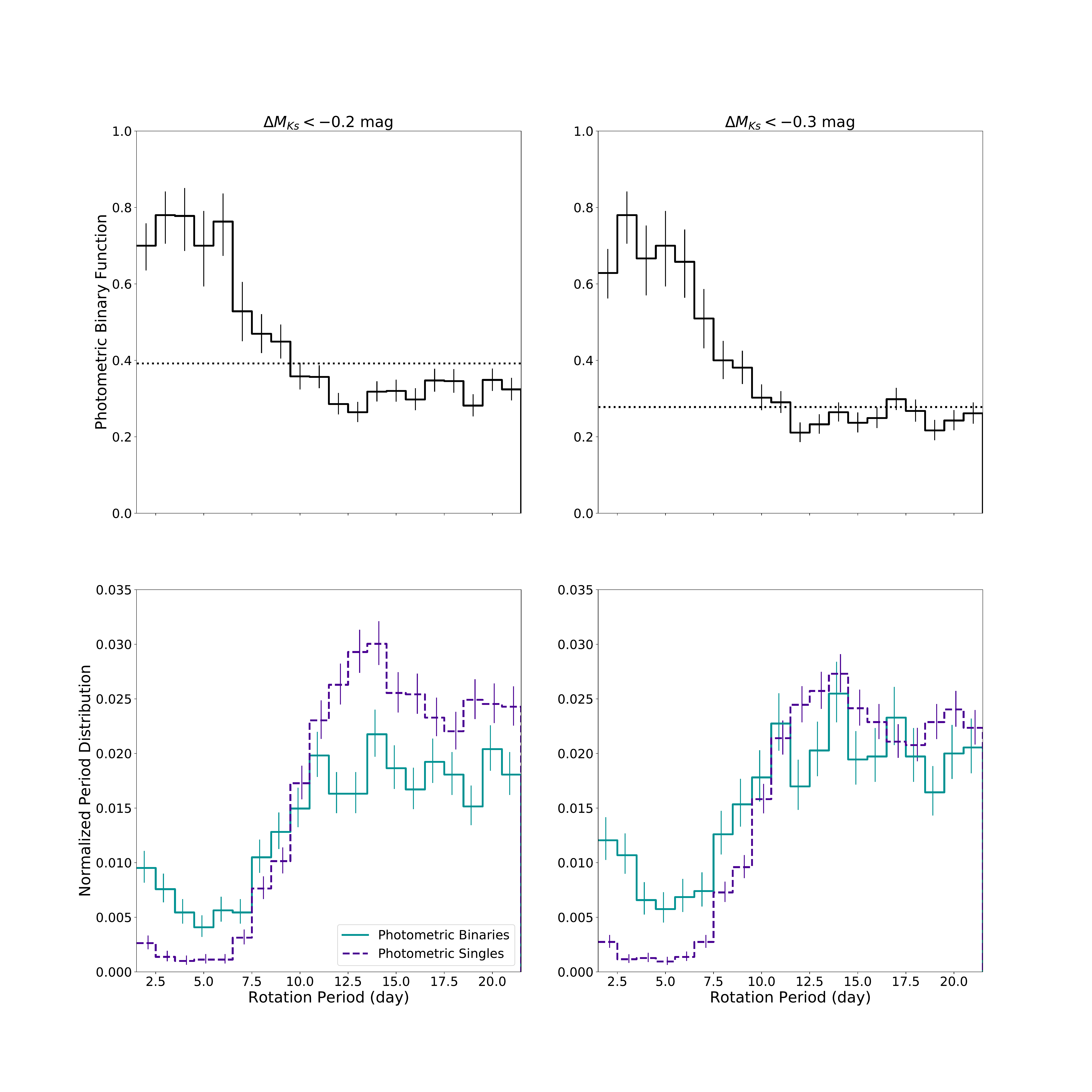}
    \caption{\emph{Top Left:} The photometric binary function for the 
        inclusive photometric binary threshold (\(\Delta \MK{} < -0.2\) mag). The 
        photometric binary fraction for the full sample analyzed by 
        \citet{McQuillan14} is shown as the dotted line. Error bars are 
        1-\(\sigma\) binomial confidence intervals. \emph{Top Right:} Same as 
        the previous plot, except using the more conservative threshold 
        (\(\Delta \MK{} < -0.3\) mag). \emph{Bottom left:} The period distribution of the photometric binary (solid green) 
        and photometric single (dashed purple) samples, each normalized by the 
        total number of photometric binaries and photometric single stars 
        analyzed by \citet{McQuillan14}, respectively, using the inclusive
        photometric binary threshold. Error bars are 1-\(\sigma\) Poisson 
        confidence intervals. \emph{Bottom Right:}
        Same as the previous plot, except with the more conservative vertical
        displacement threshold.}\label{fig:binary_fraction}
\end{figure*}

In order to understand the low-period edge of the single-star population, as
well as the cutoff period for the tidally-synchronized binaries, we need to
quantify the behavior illustrated in
\cref{fig:mcq_rapid_excess,fig:mcquillan_transition}. While we proceed
using only the photometric sample to derive statistics, we note that repeating
the analysis with the much smaller spectroscopic sample yields consistent, but
less constrained results.

One measure to trace
changes in behavior between the binaries and single stars with rotation is 
what we call the ``photometric binary function'', which denotes the 
photometric binary fraction as a function of rotation period. The photometric 
binary function is shown in the top row of \cref{fig:binary_fraction}. As was 
qualitatively shown in \cref{fig:mcquillan_transition}, the photometric binary
function is high for rotation periods below 7 days, above which it begins to
decrease, which is expected for the contribution of the
single-star tail. However, the photometric binary function does not reach the 
mean binarity of the full sample until around 9--11 days, after which the 
single-stars dominate the population.

To quantify the significance of the excess binary fraction in the rapid rotators, 
we compare the photometric binary fraction of the short-period bin to the
binary fraction of the full sample. As mentioned previously, we include the 
\Kepler{} eclipsing binaries, to accurately characterize the binarity of
short-period systems.

We also find that the period dependence of the photometric binaries and
photometric singles differ in the short-period domain. The period distribution
of photometric binaries as well as the period distribution of photometric 
single stars is given in the bottom row of \cref{fig:binary_fraction} as a 
function of period for both the inclusive and 
conservative photometric binary thresholds. The number of photometric binaries 
and photometric single stars both decrease in the short-period regime compared to the 
long-period regime; however the photometric single stars decrease much more 
steeply than the photometric binary stars. The sudden drop in photometric 
single stars implies a mixed photometric binary/single population at long 
periods, and a binary-dominated population at short periods. As seen with the
photometric binaries, a tail of single stars appears at rotation periods 
greater than 7 days for both the inclusive and conservative photometric binary 
thresholds. The relative contributions of photometric and single stars then 
become fixed past 11 days, consistent with the flattening of the photometric 
binary fraction. 

\subsection{Comparison with Eclipsing Binaries}
\label{sec:synch}

Given the radical difference in behavior between the photometric binaries and
photometric single stars between the short- and long-period regimes, we propose that the 
rapid rotators are dominated by tidally-synchronized binaries. The 
short-period systems that do not show significant vertical displacement are 
likely low mass-ratio systems. This interpretation of the photometric singles 
in the rapid rotators is supported by comparing the photometric binary 
fraction of the eclipsing binaries to that of the rapid rotators. The 
photometric binary fraction of the eclipsing binaries with orbital periods 
between 1.5--7 days is \(83^{+7}_{-10}\%\) with the inclusive threshold and 
\(76^{+8}_{-11}\%\) with the conservative threshold, while the photometric 
binary fraction of the rapid rotators with rotation periods between 1.5--7
days is \(67^{+3.3}_{-3.5}\%\) with the inclusive threshold and
\(59^{+3.6}_{-3.6}\%\) with the conservative threshold.  Given the
uncertainties, the two photometric binary fractions are just outside of one 
sigma away from each other. A slightly higher photometric binary fraction for
the eclipsing binaries is also to be expected because photometric binaries 
have larger secondaries, hence have a larger eclipse probability.

\begin{figure}[htb]
    \centering
    \plotone{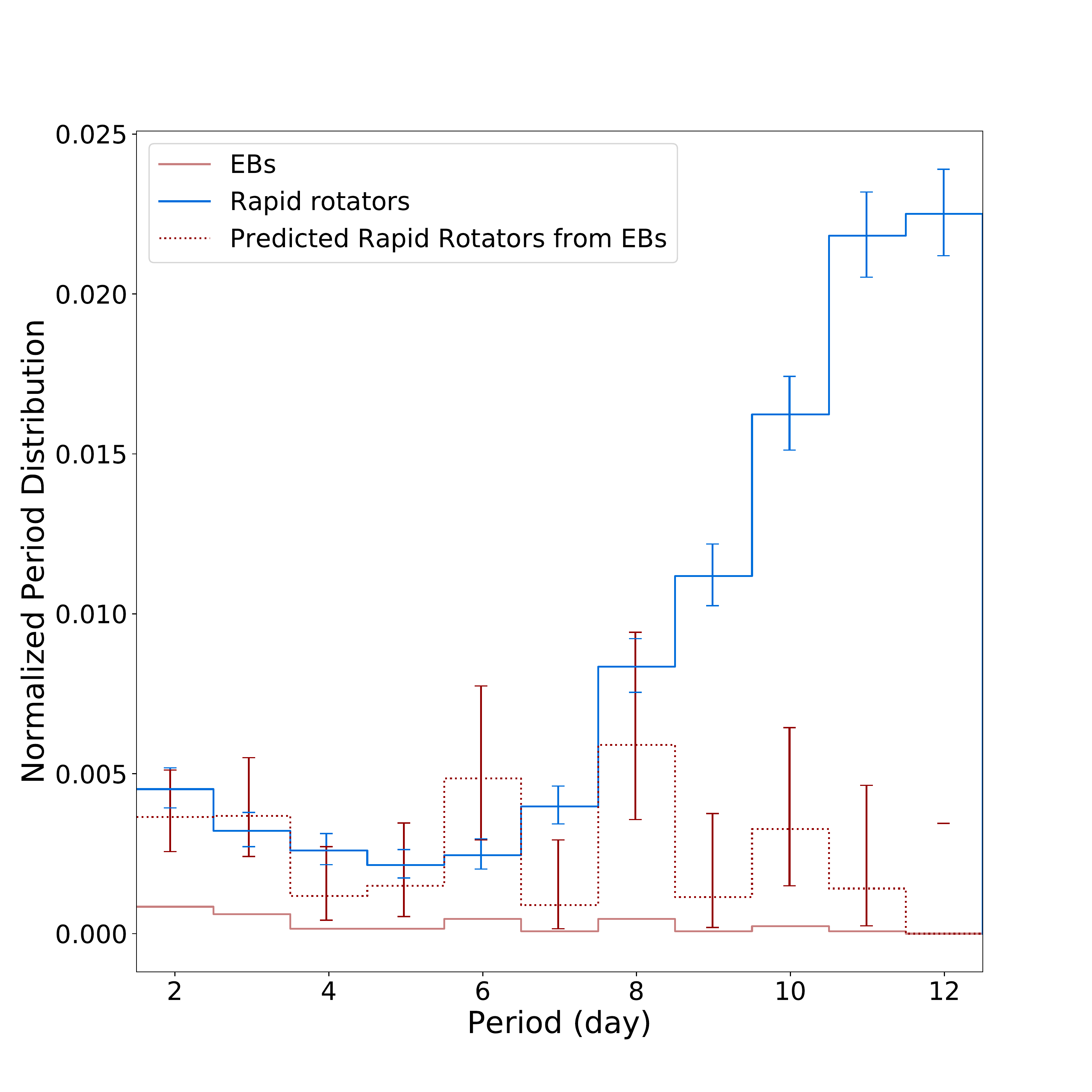}
    \caption{Comparison of the rapid rotator and eclipsing
    binary period distributions. The period distribution of the 
    \citet{McQuillan14} sample is shown as a blue solid histogram, while the 
    period distribution of the eclipsing binaries is shown as the red 
    histogram. Both distributions are normalized to the total number of 
    objects analyzed by \citet{McQuillan14} and in the EB catalog. The 
    predicted EB distribution (red dashed line) uses the observed EB orbital
    period distribution to predict the period distribution of binaries 
    inclined enough to show starspot modulation, but not inclined enough to 
    eclipse. Error bars represent 1-\(\sigma\) Poisson confidence 
intervals.}\label{fig:eclipseprob}
\end{figure}

The rotation period distribution of rapid rotators also agrees well with the 
expected orbital period distribution of a population of non-eclipsing
binaries, providing further evidence that the rotation period and orbital
periods are identical. We demonstrate this finding by comparing the observed
rotation sample to a predicted population of non-eclipsing binaries derived
using the eclipsing binary sample and the geometrical corrections in
Section~\ref{sec:ebmodel}.
In addition to the corrections from eclipse probabilities, we correct for the
detectability of rotation periods caused by inclination. K2 observations of the 
Pleiades found that 8\% of the sample did not show variations due to starspots 
\citep{Rebull17}, which is similar to the fraction of objects previously found 
to have inclinations too high to show starspots \citep{Jackson10}. Since 
rotation periods in the Pleiades are on the order of several days, comparable to 
the rotation periods in this sample, we treat 8\% of the total sample as being 
too highly-inclined to have observable starspot periods. Given these assumptions, 
and the observed population of eclipsing binaries, the predicted orbital period 
distribution of non-eclipsing binaries in \cref{fig:eclipseprob}. 

The orbital and rotation period distributions match well up to 7 days, indicating
that the rotational period distribution reflects the orbital period
distribution for the eclipsing binaries. The period at which the data diverges
from the prediction is also similar to the period where the binary fraction 
drops in \cref{fig:binary_fraction}. The concordance of these two periods 
suggests that the short-period systems follow a rotational period distribution 
compatible with the orbital period distribution of eclipsing binaries, while 
the long-period systems follow a different rotational
period distribution. Our physical interpretation for this phenomenon is that 
the short-period systems are tidally-synchronized binaries, which remain 
synchronized at least through an orbital period of 7 days. At periods longer
than 7 days, the picture becomes more complicated, with the transition from
synchronized to unsynchronized systems overlapping with the tail of
rapidly-rotating single stars and wide binaries. Determining the
synchronization cutoff period will require disentangling these populations.

Assuming that the rapid rotators and eclipsing binaries are two disjoint 
samples of the same population between 1.5--7 periods, we can combine both
samples to generalize properties of the rotation sample studied by
\citet{McQuillan14}, assuming that they would have correctly identified the
rotation period of the eclipsing binary as the orbital period. We calculate 
that \(1.9 \pm 0.1\%\) of the sample analyzed by \citep{McQuillan14} have
rotation periods between 1.5--7 days. As noted previously, the
rapid rotators and eclipsing binaries should be sensitive to 92\% of the most
inclined systems. Correcting for the low-inclination systems yields that \(2.0
\pm 0.1\%\) of systems analyzed by \citet{McQuillan14} should be synchronized
binaries with orbital periods between 1.5--7 days. Therefore, synchronized
binaries are expected to be present in rotation samples at that 
level. Future binary surveys may use photometric binaries to select systems
appropriate for further study. In a sample of pure photometric binaries, the 
short-period systems will be more prominent; systems with rotation periods
between 1.5--7 days make up \(3.4 \pm 0.3\%\) of photometric
binaries under our inclusive threshold, and \(4.5 \pm 0.4\%\) under our
conservative threshold.

\subsection{The High Photometric Binary Fraction}

\begin{figure}[htb]
    \centering
    \plotone{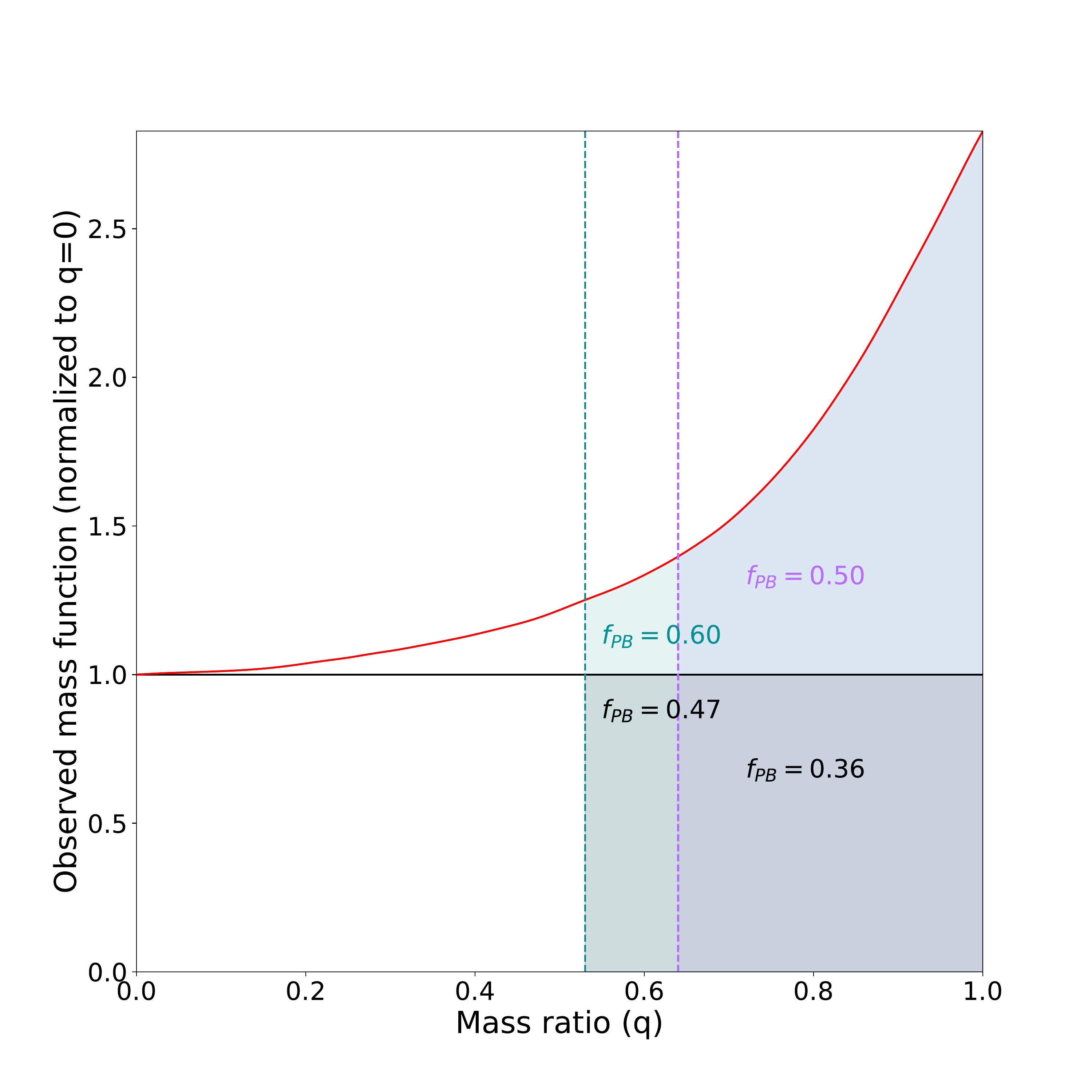}
    \caption{Observed mass distribution for a magnitude-limited binary population with a
flat mass function and constant spatial density. Without considering the
Malmquist bias, the flat mass distribution is given as the horizontal black
line. The fraction of photometric binaries for the flat mass function is 36\%
for the conservative threshold and 47\% for the inclusive threshold. The red
line denotes the observed mass distribution taking into account that higher
mass-ratio binaries are seen out to larger distances. Taking into account the
Malmquist bias raises the photometric binary fraction to 50\% for the
conservative threshold and 60\% for the inclusive
threshold.}\label{fig:malmquist}
\end{figure}

With evidence that the short-period systems are dominated by binaries, we
can compare the observed photometric binary fraction to that expected from a
flat mass-ratio distribution. As noted in the previous section, the observed photometric
binary fraction for the rapid rotator sample is 59\% using the conservative
threshold and 67\% using the inclusive threshold. If we assume a flat
mass-ratio distribution, as shown in \cref{fig:malmquist}, we expect the 
photometric binary fraction to equal the fraction of systems with mass-ratio 
large enough to exceed the given vertical displacement threshold, denoted as
the dark shaded region. Based on the model presented in \cref{fig:binary_teff}, 
the mass-ratio limits for our thresholds according to MIST models are 0.64 for 
the conservative threshold and 0.53 for the inclusive threshold. These result in
expected photometric binary fractions of 36\% and 47\% for the conservative and
inclusive thresholds, respectively. Both are substantially smaller than the
observed fractions. 

One factor which could raise the observed binary fraction is Malmquist bias. 
While the \Kepler{} selection function is heterogeneous and complex, distances
were not available for the bulk of the sample at the time. As a result, 
photometric binaries, which are intrinsically more luminous than single stars, 
would be detected out to a larger volume. 

We illustrate the effect of Malmquist bias for
an artificial population with constant spatial density and a flat mass
function in the right panel of \cref{fig:malmquist}. As expected, the observed
mass function increases because the photometric binaries are detected in larger
volumes. Using the same minimum mass-ratio as with the flat mass-ratio
distribution, including the Malmquist bias increases the photometric binary 
fraction from 37\% to 50\%, which is much closer, but still smaller than the 
observed photometric binary fraction by 2-\(\sigma\). We note that this 
treatment of Malmquist bias is an upper limit to the effect because the true 
spatial density of the \Kepler{} field should encounter an expoential dropoff, 
which would reduce the number of distant photometric binaries.  

The tension between the predicted and observed photometric binary
fractions still exists even when the most common selection effects are taken into
account. It's likely that resolving this will requires either that the 
mass-ratio for synchronized binaries is tilted toward higher-mass systems, that 
there are more subtle selection effects in the \Kepler{} sample, or that
there are higher-order systems which aren't accounted for in our simple toy
models. Attempting a detailed reconstruction of the intrinsic 
mass-ratio distribution of tidally-synchronized binaries is beyond the scope 
of this analysis.

\subsection{The Synchronization Threshold}

The 7 day limit that we get for the synchronization period cutoff is compatible
with previous theoretical expectations. The original formulation by
\citet{Zahn77} included an expression for the synchronization timescale for a
binary system given as \(t_{sync} = 10^4 ((1+q)/2q)^2 P^4 / R^6\) where \(t_{sync}\) is
in years, \(q\) is the mass-ratio, \(P\) is in days, and \(R\) is the radius in
solar radii. For a 4 Gyr old equal-mass field binary, this
corresponds to a synchronization period of 15 days, which is well within the
constraints. \citet{Claret97} revisited the formulation of \citet{Zahn77} and
instead advocated substituting a different expression for the apsidal motion 
constant which made tidal dissipation on the main sequence a factor of 5
smaller. The new expression corresponds to a synchronization period of 10 days,
which is also consistent with our findings.

From this calculation, it's clear that the 7-day limit we observe is caused by
the age distribution of the \Kepler{} field as opposed to the tidal properties
of the binaries. In order to probe the synchronization threshold for
close binaries, rotation periods will be needed for a substantially older
population than the \Kepler{} field. This may be achieved through a \Kepler-like
survey for a specifically older stellar population, or by filtering out young
stars through kinematics.

\section{Conclusion}
\label{sec:conclusions}

We calculated \(Ks\)-band luminosity excesses above the main sequence for the full
\citet{McQuillan14} sample. As seen in both a spectroscopically-characterized
subset as well as in the general sample, rapid rotators with periods between
1.5--7 days show substantially larger luminosity excesses than the slower
rotators. At periods longer than 7 days, the rapidly rotating tail of the
single-star population grows more numerous before dominating the sample at
periods longer than 9 days. 

We propose that the excess of binaries among rapid rotators is a sign that
tidally-synchronized binaries dominate the short-period end of the rotation
distribution. When comparing the rotational period distribution of the rapid 
rotators to the orbital period distribution of \Kepler{} eclipsing binaries, we
find that the rapid rotators are consistent with being synchronized, 
non-eclipsing binary systems showing starspot modulation, up to periods of 7
days.

Our hypothesis that the rapid rotators are tidally-synchronized systems can be
readily tested using time-resolved moderate-resolution spectroscopy. A binary
with two 0.7 M\(_\sun\) stars in a 7 day period orbit should have an RV
semiamplitude of 62 \kms. If these systems were truly synchronized, the
RV-derived orbital period should be identical to the photometric rotation
period. A list of the rapid rotators with 1 day < \(P_{rot}\) < 7 day, along
with relevant information is shown in \cref{tab:rapidrot}.

Assuming that the rapid rotators are tidally-synchronized systems, we find a
lower limit to the synchronization period of 7 days imposed by the rapid tail
of the population experiencing single-star angular momentum evolution, which is
consistent with most current theories of stellar tides. To derive a more
stringent upper limit, either the rapidly-rotating tail of the single-star
distribution needs to be modeled and subtracted, or stars undergoing 
single-star angular momentum evolution need to be excluded. The latter can be
done with either radial velocity monitoring to measure radial velocity
variability and to confirm the orbital period, or by measuring rotation periods
for the eclipsing binaries, as done by \citet{Lurie17}. The list of 629
objects in \cref{tab:marginalrot} with rotation periods between 7--11 days, 
which bracket the synchronization cutoff periods found in previous studies,
make a natural base sample for follow-up studies to better constrain the 
cutoff period for cool \Kepler{} dwarfs.

RV follow-up of our non-eclipsing close binaries will provide a good resource
for understanding these systems. RV orbital periods for both eclipsing and 
non-eclipsing systems 
will provide additional information on the subsynchronous binary population 
discovered in \citet{Lurie17}. The rapid rotators would also make an ideal 
sample for studies of the mass-ratio distribution for close binaries. Because 
they were detected through rotation, they should only be weakly biased with 
mass-ratio, which has long been a concern with studies of stellar multiplicity
\citep{Halbwachs03}.

A background population of tidally-synchronized binaries requires caution for
interpreting gyrochronology in the \Kepler{} field. Attempting to characterize
the age distribution of the \Kepler{} field through gyrochronology without 
taking synchronized binaries into account would under-estimate the age, 
potentially biasing future population studies of \Kepler{} stars or planets. 
In order to successfully short-period systems into a gyrochronology, either 
age diagnostics independent of rotation or activity would have to be used, or
tidally-synchronized binaries would have to be individually excluded, or
modeled as a population.

We emphasize that a tidally-synchronized background is not unique to the
\Kepler{} field. All studies of rotation, including previous work in clusters,
will eventually have to incorporate tidally-synchronized binaries to
successfully calibrate gyrochronology and angular momentum evolution models.
The \Kepler{} field, which is rich in synchronized binaries, is an excellent 
source of additional systems to study.

We also note that the \(Ks\)-band bolometric corrections in the MIST 
isochrones overpredict the dependence on metallicity, even near solar 
metallicity. We derived corrections for trends imposed by the MIST isochrones 
which reduced the scatter in the single-star sequence. In the era of \Gaia{} 
parallaxes, additional testing can be done to validate bolometric corrections 
in the sparsely-calibrated NIR region.

\acknowledgments

G.S, M.P. and D.T acknowledge support from NASA ADP Grant NNX15AF13G and from
the National Sscience Foundation via grant AST-1411685 to The Ohio State
University. We also thank the anonymous referee whose comments substantially
improved the clarity of the manuscript.
Funding for the Sloan Digital Sky Survey IV has been provided by the Alfred P.
Sloan Foundation, the U.S. Department of Energy Office of Science, and the
Participating Institutions. SDSS-IV acknowledges support and resources from the
Center for High-Performance Computing at the University of Utah. The SDSS web
site is www.sdss.org. SDSS-IV is managed by the Astrophysical Research
Consortium for the Participating Institutions of the SDSS Collaboration
including the Brazilian Participation Group, the Carnegie Institution for
Science, Carnegie Mellon University, the Chilean Participation Group, the
French Participation Group, Harvard-Smithsonian Center for Astrophysics,
Instituto de Astrof\'isica de Canarias, The Johns Hopkins University, Kavli
Institute for the Physics and Mathematics of the Universe (IPMU) / University
of Tokyo, Lawrence Berkeley National Laboratory, Leibniz Institut f\"ur
Astrophysik Potsdam (AIP), Max-Planck-Institut f\"ur Astronomie (MPIA
Heidelberg), Max-Planck-Institut f\"ur Astrophysik (MPA Garching),
Max-Planck-Institut f\"ur Extraterrestrische Physik (MPE), National
Astronomical Observatories of China, New Mexico State University, New York
University, University of Notre Dame, Observat\'ario Nacional / MCTI, The Ohio
State University, Pennsylvania State University, Shanghai Astronomical
Observatory, United Kingdom Participation Group, Universidad Nacional
Aut\'onoma de M\'exico, University of Arizona, University of Colorado Boulder,
University of Oxford, University of Portsmouth, University of Utah, University
of Virginia, University of Washington, University of Wisconsin, Vanderbilt
University, and Yale University.  This publication makes use of data products
from the Two Micron All Sky Survey, which is a joint project of the University
of Massachusetts and the Infrared Processing and Analysis Center/California
Institute of Technology, funded by the National Aeronautics and Space
Administration and the National Science Foundation. 

\facility{Kepler, Gaia, CTIO:2MASS, ARC}

\software{MIST \citep{Choi16}, Astropy \citep{astropy}, IPython \citep{PER-GRA:2007}, Scipy
\citep{jones_scipy_2001}, NumPy \citep{van2011numpy}, Matplotlib \citep{Hunter:2007} }

\onecolumngrid
\begin{deluxetable}{llcccccc}
\tablecaption{\Kepler{} Rapid Rotators\label{tab:rapidrot}}
\tablehead{\colhead{KIC} & \colhead{APOGEE ID} & \colhead{$\Teff$} & \colhead{\(K\)} & \colhead{\MK} & \colhead{$\Delta \MK$} & \colhead{$P_{\mathrm{rot}}$} & \colhead{\feh}\\ \colhead{ } & \colhead{ } & \colhead{K} & \colhead{mag} & \colhead{mag} & \colhead{mag} & \colhead{day} & \colhead{dex}}
\startdata
1570924 & 2M19232726+3707337 & 4932 & 10.625 & 3.869 & -0.323 & 3.234 & -0.08 \\
2283703 &  & 5188 & 12.878 & 3.692 & -0.310 & 2.505 &  \\
2299738 &  & 5124 & 11.966 & 3.979 & -0.072 & 6.607 &  \\
2442866 &  & 4537 & 12.106 & 3.648 & -0.826 & 2.934 &  \\
2710323 &  & 5035 & 13.215 & 2.856 & -1.260 & 2.24 &
\enddata
\tablecomments{All objects in \citet{McQuillan14} with periods
        between 1.5--7 days and 4000 K < $\Teff{}$ < 5250 K. For objects with 
        APOGEE observations, their APOGEE ID and \feh{} are given. This table is 
        published in its entirety in the machine-readable format. A portion is
        shown here for guidance regarding its form and content.}
\end{deluxetable}

\begin{deluxetable}{llcccccc}
\tablecaption{\Kepler{} Synchronization Follow-up Targets\label{tab:marginalrot}}
\tablehead{\colhead{KIC} & \colhead{APOGEE ID} & \colhead{$\Teff$} & \colhead{\(K\)} & \colhead{\MK} & \colhead{$\Delta \MK$} & \colhead{$P_{\mathrm{rot}}$} & \colhead{\feh}\\ \colhead{ } & \colhead{ } & \colhead{K} & \colhead{mag} & \colhead{mag} & \colhead{mag} & \colhead{day} & \colhead{dex}}
\startdata
1296787 &  & 4803 & 12.592 & 4.196 & -0.086 & 10.002 &  \\
1571003 &  & 4965 & 12.309 & 4.245 & 0.078 & 9.726 &  \\
1722506 &  & 4504 & 11.176 & 4.219 & -0.280 & 10.653 &  \\
1724975 & 2M19290115+3715011 & 5238 & 10.412 & 3.162 & -0.797 & 10.734 & 0.00 \\
1996773 &  & 4937 & 12.716 & 4.171 & -0.017 & 8.603 &
\enddata
\tablecomments{All objects in \citet{McQuillan14} with periods
        between 7--11 days and 4000 K < $\Teff{}$ < 5250 K. For objects with 
        APOGEE observations, their APOGEE ID and \feh{} are given. This table is 
        published in its entirety in the machine-readable format. A portion is
        shown here for guidance regarding its form and content.}
\end{deluxetable}

\bibliographystyle{aasjournal}
\bibliography{references}

\end{document}